\documentclass[a4paper,11pt]{article}
\usepackage{jheppub} % for details on the use of the package, please
\usepackage{caption}
\usepackage{subcaption}
\usepackage{graphicx,array}
\usepackage{color}
\usepackage{setspace}
\usepackage{amstext}
\usepackage{amssymb}
\usepackage{slashed}
\usepackage{makecell}
\usepackage{multirow}
\usepackage{wasysym}
\usepackage{physics}
\usepackage[T1]{fontenc} % if needed
\usepackage{changepage}
\usepackage{arydshln}
\usepackage{mathrsfs} %Ralph Smith’s Formal Script Font (rsfs):Use the “mathrsfs” package. 
\RequirePackage{luatex85}
\usepackage{tikz}
\usepackage{tikz-feynman}
\tikzset{graviton/.style={decorate, decoration={snake, amplitude=.4mm, segment length=1.5mm, pre length=.5mm, post length=.5mm}, double}}
\usepackage{scalerel} %To make the size of super/subscript smaller      

\definecolor{orange}{rgb}{1,0.5,0}
\definecolor{brown}{rgb}{0.59, 0.29, 0.0}

\definecolor{note_fontcolor}{rgb}{0.80078125, 0.80078125, 0.80078125}

\definecolor{darkgreen}{rgb}{0,0.5,0}

\def\beq{\begin{equation}}
\def\eeq{\end{equation}}
\def\bea{\begin{eqnarray}}
\def\eea{\end{eqnarray}}

\def\beq{\begin{equation}}
\def\eeq#1{\label{#1}\end{equation}}
\def\eeqn{\end{equation}}
\def\beqa{\begin{eqnarray}}
\def\eeqa#1{\label{#1}\end{eqnarray}}
\def\eeqan{\end{eqnarray}}
\def\CR{\nonumber \\ }
\def\leqn#1{(\ref{#1})}

\newcommand{\centeron}[2]{{\setbox0=\hbox{#1}\setbox1=\hbox{#2}\ifdim
		\wd1>\wd0\kern.5\wd1\kern-.5\wd0\fi \copy0
		\kern-.5\wd0\kern-.5\wd1\copy1\ifdim\wd0>\wd1
		\kern.5\wd0\kern-.5\wd1\fi}}
\newcommand{\ltap}{\>\centeron{\raise.35ex\hbox{$<$}}
	{\lower.65ex\hbox{$\sim$}}\>}
\newcommand{\gtap}{\>\centeron{\raise.35ex\hbox{$>$}}
	{\lower.65ex\hbox{$\sim$}}\>}

\def\yi{y_{\rm IR}}

\makeatletter
\newcommand*{\Relbarfill@}{\arrowfill@\Relbar\Relbar\Relbar}
\newcommand*{\xeq}[2][]{\ext@arrow 0055\Relbarfill@{#1}{#2}}
\makeatother

\usepackage{cleveref}

\title{\boldmath Collider Searches for Near-Continuum Dark Matter}

\author[a]{Steven Ferrante,}
\author[a]{Lillian Luo,}
\author[a]{Maxim Perelstein,}
\author[a,b]{and Taewook Youn}

\affiliation[a]{Department of Physics, LEPP, Cornell University, Ithaca, NY 14853, USA}
\affiliation[b]{School of Physics, Korea Institute for Advanced Study, Seoul 02455, Republic of Korea}

\abstract{We study collider constraints on the near-continuum dark matter model, in which the dark sector consists of a tower of closely spaced states with weak-scale masses coupled to the Standard Model through a $Z$-portal. To capture this structure in a model-agnostic way, we introduce a minimal parameterization that encodes the dominant geometric information with three parameters. Using a custom-built Monte-Carlo tool for near-continuous spectra, we simulate DM-pair production at $\sqrt{s}=13$ TeV and subsequent cascade decays via on/off-shell $Z$ bosons, which yield events with large missing transverse momentum and high jet multiplicity. Recasting the CMS multijet$+H_T^{\rm miss}$ analysis of Run-2 data (35.9 fb$^{-1}$), we derive bounds on the model parameter space. Extrapolating these bounds, we provide High-Luminosity LHC projections (3 ab$^{-1}$). We also project sensitivities for future electron-positron colliders at $\sqrt{s}=365$ GeV and $\sqrt{s}=500$ GeV, showing substantial improvements over the HL-LHC. The current LHC sensitivity is beginning to approach the theoretically motivated region of the parameter space, while future colliders will be able to comprehensively test this model.}

\begin{document}
\maketitle
\flushbottom

\section{Introduction}
\label{sec:Intro}

The identity of dark matter (DM) remains one of the central open questions in particle physics. While a large share of experimental effort targets weakly interacting massive particles with discrete spectra, there is growing interest in dark sectors whose excitations form a continuous set of states above a finite mass threshold. Such \textit{gapped-continuum} spectra arise naturally when extra-dimensional dynamics or approximately conformal sectors are softly broken in the infrared, producing densely spaced modes whose inclusive behavior is well captured by a spectral density rather than individual resonances. This structure of \textit{Continuum Dark Matter}~\cite{Csaki:2021gfm,Csaki:2021xpy,Csaki:2022lnq} alters both production rates and decay topologies at colliders, and it weakens traditional direct-detection constraints rooted in single-particle kinematics.\footnote{For further applications of gapped continuum in particle physics and cosmology, see Refs.~\cite{Falkowski:2008fz,Stancato:2008mp,Falkowski:2008yr,Falkowski:2009uy,Bellazzini:2015cgj,Katz:2015zba,Csaki:2018kxb,Megias:2019vdb,Megias:2021mgj,Fichet:2022ixi,Chaffey:2021tmj,Aoki:2023tjm,Kumar:2018jxz,Gabadadze:2021dnk,Eroncel:2023uqf}. The gapped continuum also arise in a range of quantum field theories~\cite{McCoy:1978ta, McCoy:1978ix,Wu:1977hi,Luther:1976mt,Cabrer:2009we} and in condensed matter systems~\cite{Fradkin:1991nr, sachdev2007quantum}.}

A concrete and predictive realization employs a 5D soft-wall geometry with the Standard Model (SM) localized on a brane~\cite{Cabrer:2009we}. Regulating the soft-wall singularity\footnote{This singularity falls into the category of a ``good'' naked singularity~\cite{Gubser:2000nd}.} by inserting an infrared brane yields a \textit{near-continuum} spectrum~\cite{Ferrante2023}: a closely spaced Kaluza–Klein (KK) tower whose splittings are parametrically smaller than the gap, so that inclusive observables are accurately captured by continuum spectral densities. In this limit, sums over KK levels are traded for integrals weighted by a spectral function that encodes the bulk geometry.

Coupling the near-continuum dark sector to the SM via a Z-portal produces a well-defined collider signature. Pair-production of near-continuum states through an s-channel $Z$ is followed by multi-step cascades emitting on- or off-shell $Z$ bosons, yielding events with large missing energy and high jet/lepton multiplicities. %These qualitative features were established in a previous study~\cite{Ferrante2023} that developed a custom Monte Carlo for near-continuum kinematics and analyzed $e^+e^-$ collisions; the same mechanisms imply rich signatures at hadron colliders such as the LHC.
To describe these processes, we employ the dedicated Monte Carlo (MC) framework developed in Ref.~\cite{Ferrante2023}, which extends standard event-generation techniques to handle near-continuous spectra. General-purpose MC tools widely used in collider phenomenology, such as MadGraph~\cite{Alwall:2011uj}, are designed for discrete particle states and therefore cannot directly accommodate the continuous mass distributions that characterize near-continuum DM. The custom MC implementation allows us to generate event samples by integrating over the spectral density $\rho(\mu^2)$, preserving the correct kinematics and multiplicity patterns of the cascades.

Theoretical and phenomenological aspects of multi-component dark matter have been explored within the general Dynamical Dark Matter (DDM) framework introduced by Dienes, Thomas, and collaborators~\cite{Dienes:2012yz,Dienes:2012cf,Dienes:2013xya,Dienes:2014via,Dienes:2014bka,Boddy:2016fds,Curtin:2018ees,Dienes:2019krh,Dienes:2021cxr,Dienes:2022zbh}. The connection between the Continuum DM framework and DDM was examined in Ref.~\cite{Csaki:2021gfm}. From a collider perspective, the two frameworks share several salient features, including high-multiplicity final states~\cite{Dienes:2019krh} and multiple displaced vertices from cascade decays~\cite{Dienes:2021cxr}. In the near-continuum DM model, however, a multi-component spectrum arises naturally from a simple, well-motivated five-dimensional construction, yielding a predictive setup that enables detailed forecasts for signal characteristics.

The goal of this paper is to place LHC bounds on near-continuum DM in the Z-portal setup using Run-2 data and to project sensitivities for the HL-LHC and future lepton colliders such as the FCC-ee~\cite{FCC:2018evy,Agapov:2022bhm} and the ILC~\cite{ILC:2013jhg,ILCInternationalDevelopmentTeam:2022izu}. 
To assess collider bounds on near-continuum DM without committing to a specific model realization, we adopt a minimal phenomenological parameterization that captures the dominant geometric information through the spectral density $\rho(\mu^2)$ above the gap $\mu_0$ as a function of the invariant mass $\mu^2$. For small and large invariant masses, $\rho$ follows universal scalings inherited from soft-wall backgrounds; we glue these regimes at the intermediate ``peak" scale $\mu_p$, and normalize the spectral density with a single dimensionless factor $\rho_0$. This provides a model-agnostic envelope for LHC production and decay rates once the portal mixing is fixed.

For hadron colliders, we evaluate the resulting inclusive cross section by folding these spectral integrals with parton distribution functions. We then generate an MC sample of cascade decay events that capture the characteristic large jet multiplicity and missing transverse momentum. To estimate the current LHC bounds, we focus on the CMS multijet+$H_T^{\rm miss}$ analysis~\cite{Sirunyan_2017}, whose inclusive topology and strong sensitivity to hadronic missing energy make it particularly well-suited to this signal. The corresponding statistical inference and extrapolation to the High-Luminosity LHC (HL-LHC) are performed by scaling event yields and tracking analysis efficiencies in bins of jet multiplicity.

Near-continuum dark sectors also yield interesting signatures at future lepton colliders, where pair production proceeds through the same $Z$-portal. The clean and well-defined initial state of $e^+e^-$ machines enables precise reconstruction of the jets and missing energy observable.
Because the near-continuum states couple to the $Z$ boson analogously to wino- or higgsino-like electroweak multiplets, the resulting production cross sections remain sizable even at moderate center-of-mass energies. As a result, lepton colliders can achieve significantly better sensitivity than the LHC despite their lower energy reach. This makes the near-continuum scenario a particularly promising physics case for future facilities such as the ILC or FCC-ee.

The paper is organized as follows. In Section~\ref{sec:model} we review the near-continuum DM construction, its continuum limit, and our spectral-density parameterization. Section~\ref{sec:NCProdDecay} summarizes production and Z-portal decays relevant for hadron colliders. Section~\ref{sec:VisiblePheno} presents our LHC recast and HL-LHC projections. Section~\ref{sec:eeProj} outlines prospects at future lepton colliders. We conclude in Section~\ref{sec:conc}.

\section{Model}	
\label{sec:model}

In this section, we briefly review the explicit 5D model of near-continuum dark matter, introduced in~\cite{Csaki:2021gfm, Csaki:2021xpy, Ferrante2023}. We then describe the parametrization of the DM spectral density used in the collider analysis of this paper.  

\subsection{5D Model of Continuum DM} 
\label{sec:5Dcont}

Following Ref.~\cite{Cabrer:2009we}, we start with a scalar field coupled to gravity in a 5D space ($x^{\mu}, y$) with metric $ds^{2} = e^{-2A(y)}\eta_{\mu\nu}dx^{\mu}dx^{\nu}-dy^{2}$ and a $\mathbb{Z}_{2}$ orbifold symmetry under which $y\rightarrow -y$ and the scalar field and the metric are even. The action for this coupled system is 
\begin{align}
    S = \int d^{5}x 
    \sqrt{-g}
    ( M_{5}^{3}R - 3(\partial\phi)^{2} - V(\phi) )
    - 
    \int d^{4}x
    \sqrt{-g_{\text{ind}}} \, \lambda(\phi)\big|_{y=0},
\label{eq:softwallaction}
\end{align}
where $M_{5}$ is the 5D Planck scale, and $V(\phi)$ and $\lambda(\phi)$ are the bulk and brane contributions to the scalar potential. It was described in~\cite{Cabrer:2009we} that these potentials can be written in terms of a ``superpotential" $W(\phi)$, where $V = 3(\partial_{\phi}W)^{2} - 12W^{2}$ and $\lambda = 6W$, and that the choice
\begin{align}
    W = k(1+e^{\phi})
\label{eq:superpotential}
\end{align} 
yields the following classical solution for the graviton and scalar profiles: 
\begin{align}
    A(y) = ky - \text{log}
    \bigg(1-\frac{y}{y_{s}}\bigg), 
    \,\,\,\,\,\,\,\,
    \phi(y) = -\text{log}\, k(y-y_{s}).
\label{eq:softwallgeometry}
\end{align}
This solution predicts a curvature singularity at the point $y_{s}$. The interpretation is that spacetime ends here and this is what is referred to as the `soft wall', which is schematically shown in Figure \ref{fig:NearContinuumSetup}. 

In Ref.~\cite{Cabrer:2009we} it was pointed out that under 4D Kaluza-Klein (KK) decomposition, the metric in Eq.~\leqn{eq:softwallgeometry} gives rise to gapped continuum spectra for the scalar field and the graviton. The gap mass scale in both cases is given by 
\begin{align}
    \mu_{0} = \frac{3}{2}\frac{e^{-ky_{s}}}{y_{s}}.
\end{align}
The graviton spectrum also includes a zero-mode, so that 4D Newtonian gravity is reproduced at large distances. 

The Standard Model (SM) is incorporated in this construction by adding a 4D brane at $y=R < y_{s}$ on which the SM fields are localized, as depicted in Figure \ref{fig:NearContinuumSetup}. We take $k R \sim \mathcal{O}(10)$, which preserves the Randall-Sundrum (RS) solution to the hierarchy problem \cite{Randall:1999ee}, where mass scales at $y=R$ are warped down from the scale of gravity $\sim M_{5}$  by the factor $e^{-kR}$. We also assume there is negligible back-reaction on the geometry (\ref{eq:softwallgeometry}), i.e. the SM brane tension is small. 

In~\cite{Csaki:2021gfm} it was proposed that dark matter could be introduced as a 5D field propagating on the background of the classical geometry defined in Eq.~\leqn{eq:softwallgeometry}. The minimal model is a scalar field $\Phi$ with a stabilizing $\mathbb{Z}_{2}$ symmetry, under which $\Phi$ is odd. Its bulk action is 
\begin{align}
    S = \int 
    d^{5}x
    \sqrt{-g}
    \left( g^{MN}\partial_{M}\Phi\partial_{N}\Phi 
    - m^{2}\Phi^{2} \right),
\end{align}
where $\Phi$ is assumed to propagate only between the SM brane and the singularity. Under 4D KK decomposition, $\Phi$ also has a gapped continuum spectrum with the same gap scale $\mu_0$. These continuum excitations are identified as the DM.

\begin{figure}[t]
\centering
\includegraphics[width=6in]{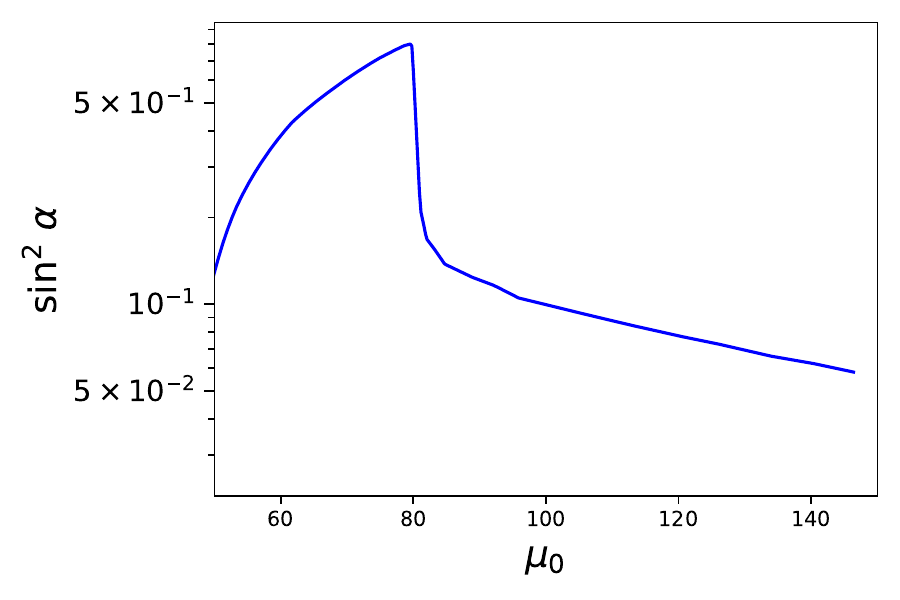}
\caption{Mixing angle $\sin^{2}\alpha$ required to reproduce the observed DM relic abundance, as a function of the gap scale $\mu_{0}$. From Ref.~\cite{Csaki:2021gfm}.}
\label{fig:dmrelic}
\end{figure}

A minimal way of coupling dark matter to the SM in this setup is the ``Z-portal" model proposed in~\cite{Csaki:2021gfm,Csaki:2021xpy}: 
\begin{align}
    S = \int d^{4}x 
    \sqrt{-g_{\text{ind}}}\,\left(
    g^{MN}D_{M}\chi^{\dagger}D_{N}\chi 
    - m_{\chi}^{2}|\chi|^{2}
    + \hat{\lambda}k^{1/2}\Phi(R)\chi H + h.c. \right)\,,
\end{align}
where $H$ is the SM Higgs doublet, and $\chi$ is an additional electroweak-doublet scalar field localized on the SM brane and odd under the $\mathbb{Z}_{2}$. Upon electroweak symmetry breaking, integrating out the neutral component of $\chi$ gives rise to interactions between $\Phi$ and the SM $W$ and $Z$ gauge bosons with effective coupling 
\begin{align}
    g_{eff} = g_{Z}\,\text{sin}^{2}\alpha,
\end{align}
where $g_{Z} = \sqrt{g^{2} + g'^{2}}$ is the SM $Z$ coupling, and the mixing angle $\sin^{2}\alpha$ is typically between 0.01 and 0.1 to reproduce the observed DM relic abundance. In Figure~\ref{fig:dmrelic}, we show the value of $\sin^{2}\alpha$ as a function of the gap scale $\mu_{0}$ as calculated in~\cite{Csaki:2021gfm}.  For small gap scale $\mu_{0} < m_{W}$, the dominant contribution to the freeze-out is $\phi\phi \rightarrow f\bar{f}$ through an s-channel $Z$ boson. For $\frac{m_{Z}}{2}\ll \mu_{0}<m_{Z}$, the thermally averaged cross section for this process scales like $\mu_{0}^{-2}$, so $\sin^{2}\alpha$ increases with larger $\mu_{0}$ for a fixed relic abundance.  After the $W$ threshold $\mu_{0} \gg m_{W}$, processes involving on-shell vector boson production can also contribute to the annihilation rate, such as $\phi\phi\rightarrow W^{+}W^{-}$  and $\phi\phi\rightarrow ZZ$. The thermally averaged cross sections of both processes\footnote{The process $\phi\phi \to W^{+}W^{-}$ can proceed either via a contact interaction or through an $s$-channel $Z$ boson. The corresponding cross sections scale as $\mu_{0}^{2}$ and $\mu_{0}^{-2}$, respectively, so only the contact interaction remains relevant at large $\mu_{0}$.} scale like $\mu_{0}^{2}$, indicating that the mixing angle drops for larger gap scale to keep the same relic abundance~\cite{Csaki:2021gfm}. 

\subsection{Near-Continuum DM}

\begin{figure}
\centering
\includegraphics[width=4.6in]{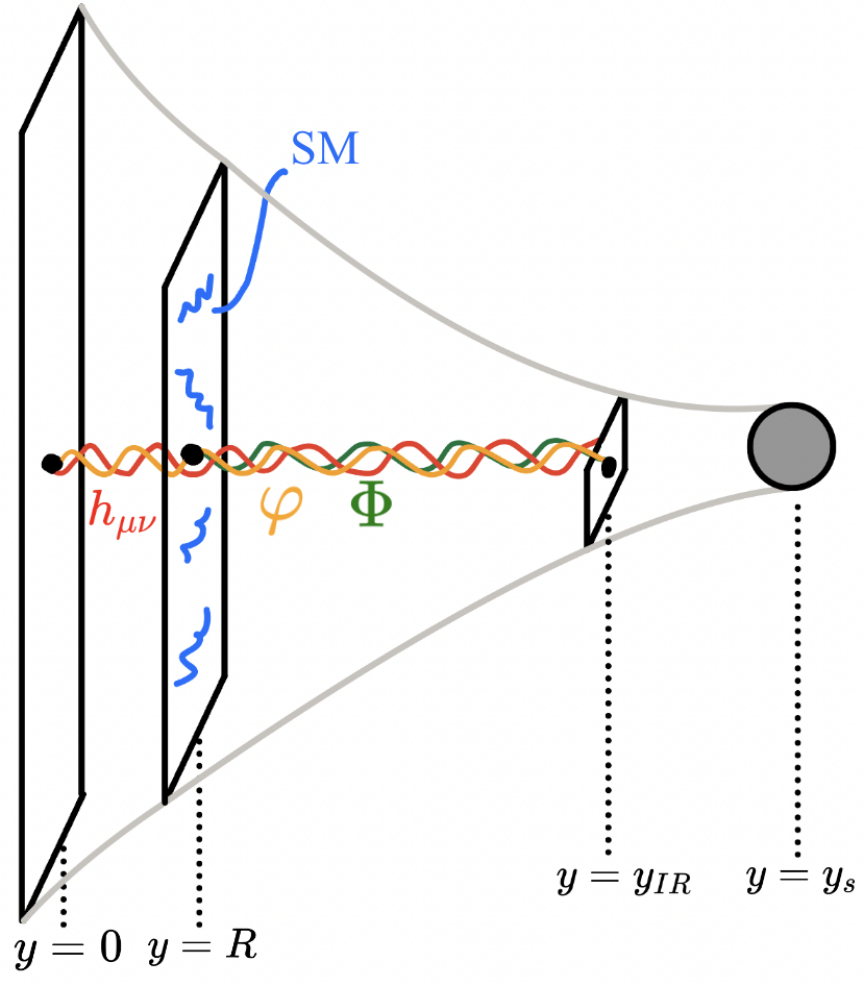}
\caption{5D setup of the near-continuum dark matter model. From Ref.~\cite{Ferrante2023}.}
\label{fig:NearContinuumSetup}
\end{figure}

To describe the physics near the singularity $y_{s}$, a proper theory of quantum gravity is needed for curvatures that exceed the 5D Planck scale. Such a description is expected to provide an effective cutoff, avoiding large curvatures. In Ref.~\cite{Ferrante2023}, this was modeled by introducing an infrared (IR) regulator brane near the singularity, at $y_{\text{IR}} \lesssim y_{s}$. With the IR regulator brane, the space ends at $y_{\text{IR}} $ rather than $y_{s}$, excluding the problematic singular region near the singularity. To achieve this the action for the regulator brane must take the form 
\beq
    S_{\text{IR}-\text{reg}} = 
    -\int d^{4}x
    \sqrt{-g_{\text{ind}}} \, \gamma(\phi)\big|_{y=y_{\text{IR}}},
\eeq{Sreg}
where $\gamma = -\lambda = -6W$. This regulator brane is also schematically depicted in Figure~\ref{fig:NearContinuumSetup}. In the presence of the regulator brane, KK decomposition of scalar or graviton bulk fields will result in discrete KK towers rather than a gapped continuum.  The characteristic mass spacing $\Delta m$ between neighboring states in the KK tower is parametrically small,
\begin{align}
    \Delta m \ll \mu_{0},
\end{align}
if the IR regulator brane is sufficiently close to the singularity: in fact, $\Delta m\to 0$ in the limit $y_{\rm IR}\to y_s$. In this regime, the physics of the discrete spectrum will closely resemble the physics of a true gapped continuum. For this reason we call KK spectra that satisfy this condition ``\textit{near-continuum}''. 

% \textbf{SF: Explain benchmark point? Probably not, since our `BP' is really a scan over the parameterized spectral densities of the next section ... }

The spectrum for $\Phi$ is found by substituting its KK expansion $\Phi = \sum_{n}\phi_{n}(x)f_{n}(y)$ into the bulk equations of motion that follow from the action~\leqn{eq:softwallaction}, and imposing the boundary conditions at $y=R$ and $y=y_{\rm IR}$. 
Defining $\psi_{n} = e^{-3A/2}f_{n}$ and going to conformally flat coordinates (defined by $dz = e^{A} dy$), the equation of motion for the profile is a Schrodinger-like equation:
\beq
    -\ddot{\psi}_{n} + V(z)\psi_{n} = m_{n}^{2}\psi_{n},
\eeq{Schrod}
where $\dot{()}$ denotes a derivative with respect to $z$, and $V(z) = \frac{9}{4}\dot{A}^{2}-\frac{3}{2}\ddot{A}+m^2 e^{-2A}$. For large $z$, $V(z)$ approaches a constant $\mu_{0}^{2}$, so the profiles $\psi_{n}$ are sinusoidal $\sim z_\text{IR}^{-1/2} \text{sin}(\kappa z)$, where $\kappa = \sqrt{m_{n}^{2} - \mu_{0}^{2}}$, and we see following scaling behaviors
\begin{align}
    \Delta m_{n}^{2} = m_{n+1}^{2} - m_{n}^{2} \sim \frac{m_{n}}{z_{\text{IR}}}, 
    ~~~~~~~~~
    |\psi_{n}| \sim z_{\text{IR}}^{-1/2}. 
\end{align}
In this limit, calculations of physical observables that include sums over KK modes turn into integrals over a spectral density. But how do we connect the discrete profile wavefunctions with a spectral density in this limit?  In Ref.~\cite{Ferrante2023}, it was argued that $\rho$ can be identified with a limit 
\beq
  \frac{1}{2\pi R} \, \rho(m_n^2)  \,=\, \lim_{\yi\to y_s} \frac{|f_n(R)|^2}{\Delta m_n^2},
\eeq{limita}
so that in calculations of physical observables, the sum can be turned into an integral through the replacement:
\beq
R \, \sum_n |f_n(R)|^2 \,\to\,\int \frac{d\mu^2}{2\pi}\,\rho(\mu^2).
\eeq{XS_limit}
For example, this is how inclusive cross sections for continuum KK mode production have been calculated previously~\cite{Ferrante2023, Csaki:2021gfm}.

In the near-continuum geometry, each KK mode is an independent 4D field with its own propagator that gets corrections from interactions with the SM and gravitons. The imaginary part of the loop corrections to the KK mode self-energies $\text{Im}\,\Pi_{kl}(p^{2})$ describes the decay of KK modes to lighter states. Depending on the model parameters,  $\text{Im}\,\Pi_{kl}(p^{2})$ can be larger or smaller than the mass spacing $|m_{k}^{2} - m_{k\pm 1}^{2}|$, and this determines whether the narrow width approximation (NWA) is satisfied. If NWA holds, sums over KK modes for the calculation of cross-sections and decays are performed at the level of the squared matrix element (i.e. there is no interference among the KK modes). This allows one to go beyond calculating inclusive cross sections, and perform a detailed analysis of DM cascade decays, using the familiar tools of perturbative quantum field theory.\footnote{In the opposite regime, where the NWA is not satisfied, individual KK modes cannot be used as asymptotic states in S-matrix calculations. The entire KK tower needs to be considered together as an ``unparticle", and understanding its evolution at late times is highly non-trivial. See Ref.~\cite{Megias:2023kpk} for recent work on this issue.} 

In~\cite{Ferrante2023}, the following benchmark point was considered:
\beqa
k&=&10^{10}~{\rm GeV};~~\mu_0=100~{\rm GeV};~~R^{-1}=80~{\rm GeV};~~M_5=2\cdot 10^{16}~{\rm GeV};\CR y_{\rm IR} &=&1.6\cdot 10^{-9}~ {\rm GeV^{-1}}; ~~~\sin^2\alpha\,=\,0.1
\eeqa{BM_model}
This benchmark point was found to satisfy the NWA, so calculation of collider observables of near-continuum DM was feasible. In this analysis we scan the parameter space in a neighborhood of this model (for example, we scan gap scales from $50-150$ GeV) and assume that the NWA still holds to the extent we can use the same procedure to calculate production and decay rates. In practice, we perform the scan using the phenomenological parametrization described below, rather than explicitly scanning the parameters of the 5D model. 

\subsection{Parameterization of Spectral Densities}

While the above construction provides an explicit example of continuum and near-continuum DM models, it is not unique. Any choice of classical bulk geometry which leads to the effective Schrodinger potential $V(z)$ in Eq.~\leqn{Schrod} approaching a positive constant at large $z$ will yield a gapped continuum (or, with an IR regulator brane, near-continuum) spectrum. In this work we will not attempt to build further explicit 5D models. Instead, we will use the fact that, in the near-continuum limit, the spectral density $\rho(\mu)$ is the only input from the 5D theory required by the collider analysis (up to an overall normalization from the strength of the brane-bulk coupling). We then take a phenomenological approach and consider a parametrization of spectral densities which takes into account the generic features expected in 5D constructions, but is sufficiently general to not be beholden to any specific 5D model. 

\begin{figure}
\centering
\includegraphics[width=5in]
{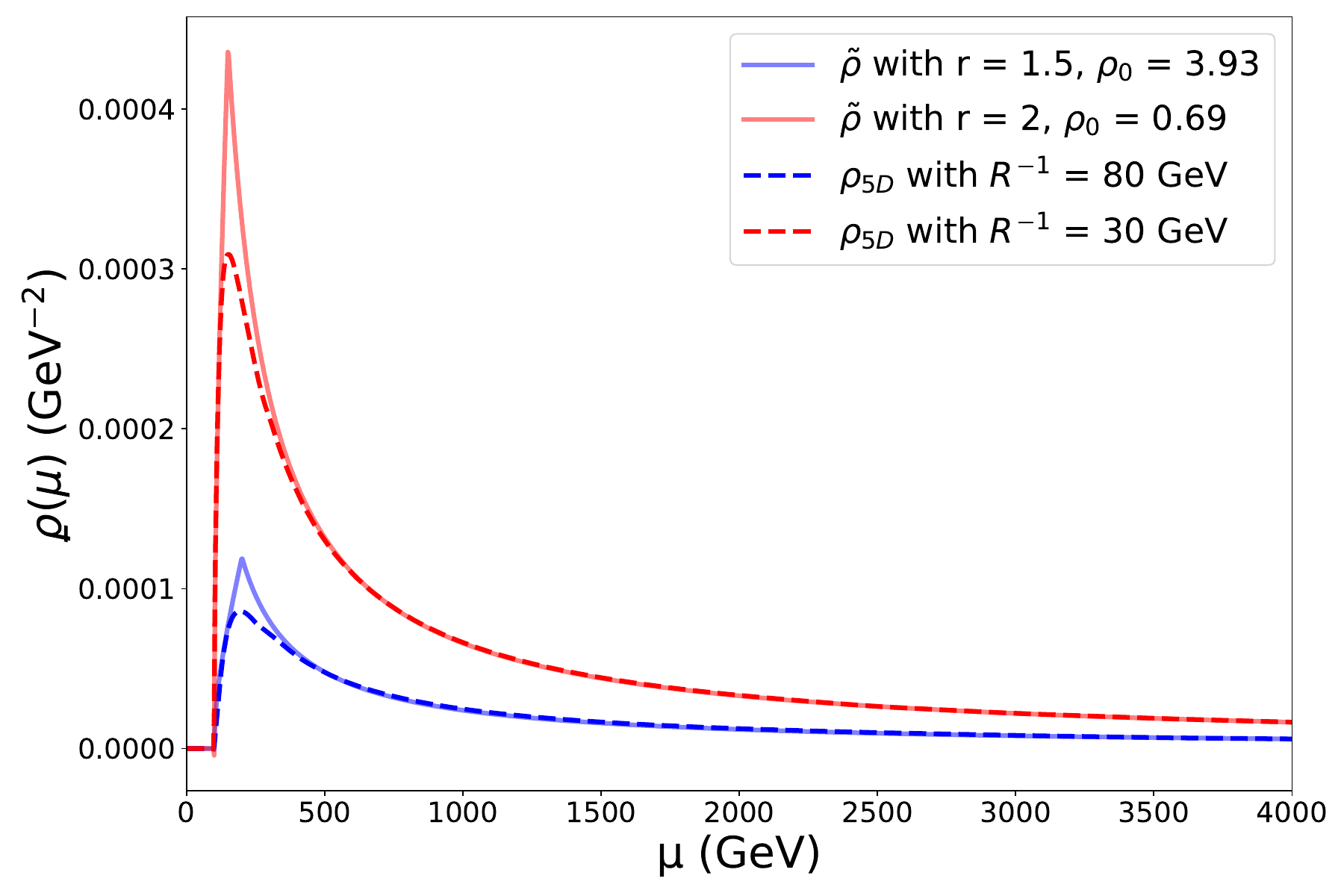}
\caption{Dashed lines: numerical results for spectral densities $\rho_{5D}$ from the explicit 5D theory described in section~\ref{sec:5Dcont}, with $\mu_{0} = 100$ GeV and SM brane located at $R^{-1} = 80$ GeV (blue) and 30 GeV (red). Solid lines: The parameterized spectral densities $\tilde{\rho}$ from Eq.~(\ref{eq:ToyRhoDef}), with $\mu_{0} = 100$ GeV and values of $r$ and $\rho_0$ chosen to match each $\rho_{5D}$.}
\label{fig:toyrhos}
\end{figure}

To model the spectral densities that can arise from different bulk geometries, we exploit the fact that for small $\mu^{2}$ they generically grow as $\rho \sim\sqrt{\mu^{2} - \mu_{0}^{2}}$ (see Appendix B of~\cite{Csaki:2021gfm}), and for large $\mu^{2}$ they fall as $\rho \sim \frac{1}{\mu}$.\footnote{The reason for the universal $1/\mu$ behavior is most readily apparent from the point of view of the 5D holographic action approach, in which the spectral density is related to the boundary-to-boundary propagator of a bulk scalar field with 4D momentum $p^2=\mu^2$. At large momentum, this propagator only probes the region of the bulk immediately adjacent to the brane, where the effects of a non-trivial 5D geometry are negligible.}
We define a generic spectral density $\tilde{\rho}$ by gluing these two functions together at a certain ``peak" mass scale $\mu_{p}$. The spectral densities are then parameterized by an overall normalization $\rho_{0}$, the gap scale $\mu_{0}$, and the dimensionless ratio $r\equiv \mu_{p}/\mu_{0}$:
\beq
    % \rho_{\mu_{0},R}(\mu^{2}) 
    \tilde{\rho}(\mu^{2})
    = 
    \begin{cases}
        \frac{\rho_{0}}{\mu_{0}^{2}}\sqrt{\big(\frac{\mu}{\mu_{0}}\big)^{2}-1} , \,\,\,\,\,\,\,\, \mu > r\mu_{0} \\
        1/(\mu_{1}\mu), \,\,\,\,\,\,\,\,\,\,\,\,\,\,\,\,\,\,\,\,\,\,\,\,\, \mu < r\mu_{0}
        % \frac{\rho_{0}R\sqrt{R^{2}-1}}{\mu_{0}}
        % \frac{1}{\mu}
    \end{cases}
\eeq{eq:ToyRhoDef}
where $\mu_{1} \equiv \frac{\mu_{0}}{\rho_{0} r\sqrt{r^{2} - 1}}$. 
For $Z$-portal continuum DM models, we expect that both $\mu_0$ and $\mu_p$ are around the weak scale. In~\cite{Csaki:2021gfm} is was also pointed out that $\rho_{0}$ is $\mathcal{O}(1)$ in phenomenologically interesting models. Examples of parametrized spectral densities for $\mu_{0} = $ 100 GeV and different values of $r$ are shown by the solid lines of Fig.~\ref{fig:toyrhos}. 

For comparison, we also plot (with dashed lines) numerical results for spectral densities from the explicit soft-wall 5D model described above\footnote{We are grateful to Seung Lee for providing the numerical solutions for spectral densities used in these plots.}, which we denote as $\rho_{\text{5D}}$. In these plots, the 5D gap scale $\mu_{0}$ was fixed at 100 GeV, and two different values of the bulk parameter $R^{-1}$, which is primarily responsible for determining the shape of the spectral density in the 5D model, were used. The parameters $\mu_0$, $r$ and $\rho_0$ of the parametrized spectral densities shown in Figure~\ref{fig:toyrhos} were chosen to match the main features of the numerical spectral densities from the 5D, such as the gap scale, the location of the maximum, and the integral $\int d\mu^2\,\rho$. As can be seen from the figure, the parametrized distributions provide a good approximation of the spectral densities predicted by the 5D model.

%. When integrated up to LHC energies (13 TeV), the integrals are $\int \tfrac{d\mu^{2}}{2\pi} \rho_{\text{5D}} \sim 270$  for $R^{-1} = 30$ GeV, and $\int \tfrac{d\mu^{2}}{2\pi} \rho_{\text{5D}} \sim 98$  for $R^{-1} = 80$ GeV. In our parameterization, setting the ratios $r$ to 1.5 and 2 for $R^{-1} = $ 30 and 80 GeV respectively, $\tilde{\rho}$ and $\rho_{\text{5D}}$ have the same integral if  $\rho_{0} \sim$  3.93 and 0.69 respectively, as shown in Figure \ref{fig:toyrhos}. 

\section{Collider Phenomenology}
\label{sec:NCProdDecay}

In this section we describe the production of near-continuum states and their subsequent decay at hadron colliders like the LHC.

\subsection{Near-Continuum DM Production}
\label{subsec:NCProduction}

\begin{figure}[t]
\centering
\includegraphics[width=5in]
{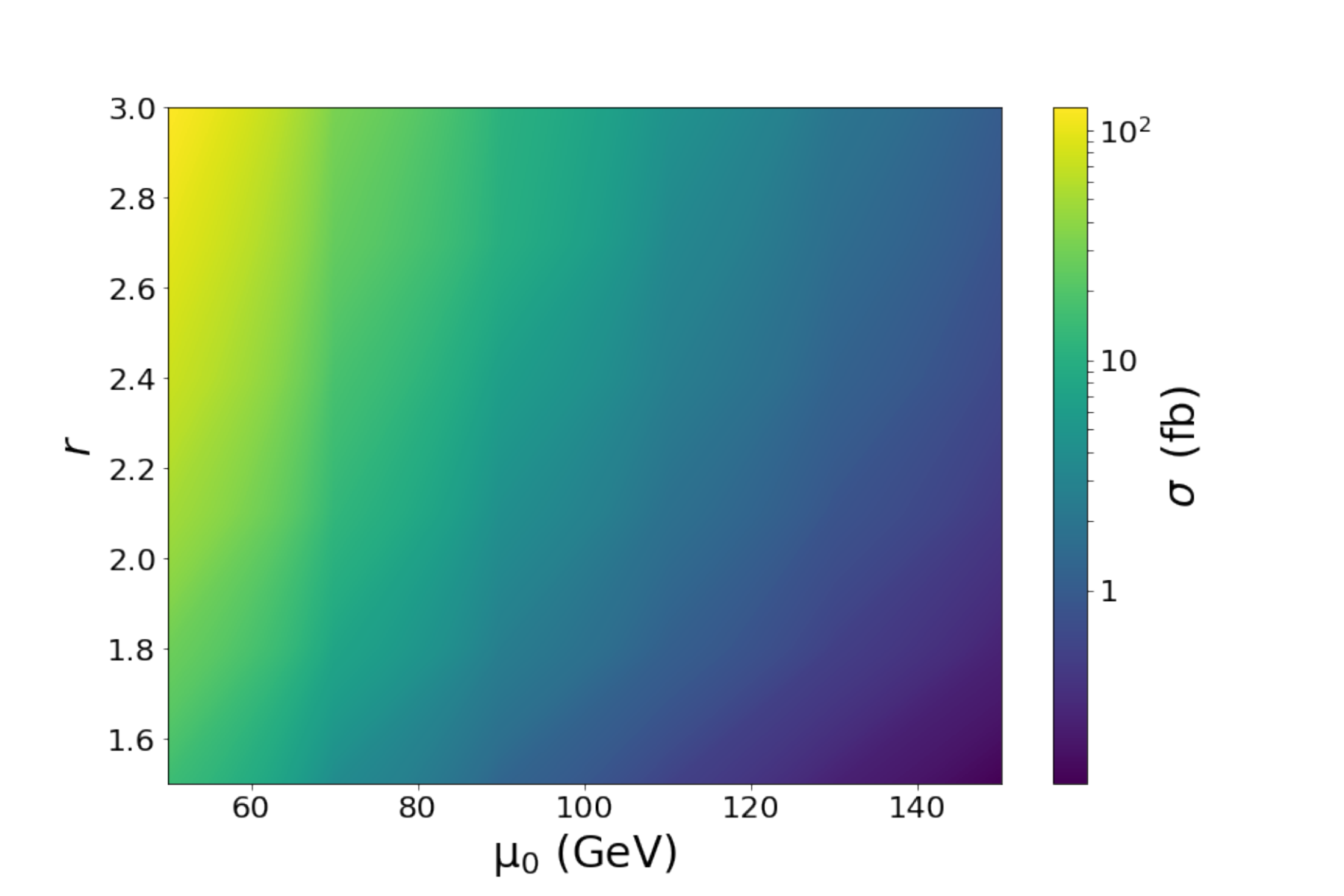}
\caption{Total cross section for pair production of near-continuum DM states in 13 TeV $pp$ collisions as a function of the gap scale $\mu_{0}$ and the ratio $r\equiv \mu_{p}/\mu_{0}$. The overall coefficients $\rho_{0}^{2}$ and $\text{sin}^{4}\alpha$ are set to 1. }
\label{fig:XSection}
\end{figure}

Near-continuum DM states are produced in the $s$-channel via their coupling to $Z$ bosons. All KK states are odd under the DM $Z_2$ symmetry, while the SM particles are even, so that the KK states must be pair-produced in SM collisions. At tree level, the only diagram is

\begin{align}
\begin{tikzpicture}
\begin{feynman}[inline=(b.base), large]
\vertex (b);
\vertex [above left =of b] (a) {\(\overline{q}\)};
\vertex [below left =of b] (c) {\(q\)}; 
\vertex [right =of b] (d); 
\vertex [above right = of d] (f1) {\(\phi(\mu)\)}; 
\vertex [below right = of d] (f2) {\(\phi(\mu')\)}; 
\diagram* {
(a) -- [anti fermion] (b) -- [anti fermion] (c), 
(b) -- [boson, edge label'=\(Z\)] (d),
(d) -- (f1),
(d) -- (f2),
};
\end{feynman}
\end{tikzpicture},
\end{align}
where $\phi(\mu)$ and $\phi(\mu')$ denote near-continuum states of distinct masses $\mu$ and $\mu'$. 
Following the same procedure as in~\cite{Ferrante2023}, we use {\tt Vegas}~\cite{Lepage:1977sw} to perform the phase space and parton distribution function (PDF) integrals in the cross section and to sample the distribution of the final state DM masses and scattering angles. The cross section can be written as the following integral 
\beq
    \sigma = 
    % \bigg( \int
    % \frac{d\sigma}{d\text{cos}\theta}
    % \bigg)
    \int
    \frac{d^{5}\sigma}
    {d\cos\theta \, d\mu d\mu' dx dy}
    ,
\eeq{}
where the integration bounds are given by 
\beq
    \int_{-1}^{1} d\text{cos}\theta
    \int_{\mu_{0}}^{\sqrt{s}-\mu_{0}}
    \frac{d\mu^{2}}{2\pi}
    \int_{\mu_{0}}^{\sqrt{s}-\mu}
    \frac{d\mu'^{2}}{2\pi}
    \int_{\frac{(\mu+\mu')^{2}}{s}}^{1} dx
    \int_{\frac{(\mu+\mu')^{2}}{xs}}^{1} dy ,
\eeq{}
and the differential cross section is
\beq
    \frac{d^{5}\sigma}
    {d\cos\theta \, d\mu d\mu' dx dy}
    = 
    (1-\text{cos}^{2}\theta)
    % \frac{g_{Z}^{2}\text{sin}^{4}\alpha }
    %    {128\pi\sqrt{s}(s-m_{Z}^{2})^{2}}
    \rho(\mu^{2}) \rho(\mu'^{2})
    \sum_{q=u,d,s} 
    c_{q}
    f_{q}(x) f_{\bar{q}}(y)
    A(\sqrt{xys}, \mu, \mu').
\eeq{eq:ProdXSec}
Here, $\theta$ is the scattering angle in the parton center-of-mass frame; $x$ and $y$ are the momentum fractions of the two incoming partons $q$ and $\bar{q}$; and the functions $f_{q}(x)$ and $f_{\bar{q}}(y)$ are the proton PDFs obtained from the CT10 global fit \cite{Lai_2010} as implemented in the {\tt Parton} Python interface, which internally uses LHAPDF \cite{Buckley_2015}. The coefficients $c_{q}$ are defined as
\beq
c_{q} = \frac{g^{2}}{2\text{cos}^{2}\theta_{w}}\sum_{i}(T^{3}_{i}-\text{sin}^{2}\theta_{w} Q_{i})^2~~~~~~~~~~~~~(i=q_{L}, q_{R}).
\eeq{eq:cq_def}
The function $A$ can be written in terms of sums and differences of the squared masses $\Sigma\equiv \mu^{2}+\mu'^{2}$ and $\Delta\equiv \mu^{2}-\mu'^{2}$: 
\beq
   A(\sqrt{s}, \mu, \mu') = 
    \frac{g_{Z}^{2}\text{sin}^{4}\alpha }
       {128\pi\sqrt{s}(s-m_{Z}^{2})^{2}}
   \bigg(
   \frac{s^{2}-2s\Sigma+\Delta^{2}}{4s}
   \bigg)^{3/2}.
   % \,(1-\cos^2\theta)
\eeq{}
Using the parameterization of the spectral densities $\tilde{\rho}$ defined in Eq.~\leqn{eq:ToyRhoDef}, this cross section is a function of $\mu_{0}$ and $r$, and has an overall factor of $\text{sin}^{4}\alpha \, \rho_{0}^{2}$.  Setting this overall factor to 1, the cross section as a function of $\mu_{0}$ and $r$ is shown in Fig.~\ref{fig:XSection}.

\subsection{Z-portal Decay}
\label{subsec:ZDecay}
After production, a DM state of mass $\mu_k>\mu_{0}$ will decay to another DM state of mass  $\mu_l\in (\mu_{0}, \mu_{k})$ via emission of a SM $Z$ boson\footnote{There is also a decay mode with an emission of a KK graviton, which is subdominant~\cite{Ferrante2023}.}: 
\beq
\phi_k \to \phi_l + Z^{(*)} \to \phi_l + f\bar{f}.
\eeq{decZ}
% \beq
% \phi_k \to \phi_l + G_m.
% \eeq{decG}
Here the $Z$ boson may be on- or off-shell depending on the DM masses, and $f$ denotes any of the kinematically accessible SM fermions. Note that the presence of a DM KK mode $\phi_l$ in the final state is required by the conserved $Z_2$ symmetry. 
% We find that for the benchmark model defined in Eq.~\leqn{BM_model}, decays into SM fermion pairs dominate.
% Following this decay, the produced DM state $\phi_l$ will itself decay into another, lighter DM state and a pair of SM fermions, and so on, resulting in a ``cascade decay" event topology. 
The decay~\leqn{decZ} is described by the diagram 
\begin{align}
\begin{tikzpicture}
\begin{feynman}
\vertex (a) {\(\phi(\mu_{k})\)};
\vertex [right=of a] (b);
\vertex [above right=of b] (f1) {\(\phi(\mu_{l})\)};
\vertex [below right=of b] (c);
\vertex [above right=of c] (f2) {\(\bar{f}\)};
\vertex [below right=of c] (f3) {\(f\)};
\diagram* {
(a) -- (b) -- (f1),
(b) -- [boson, edge label'=\(Z\)] (c),
(c) -- [anti fermion] (f2),
(c) -- [fermion] (f3),
};
\end{feynman}
\end{tikzpicture}.
\nonumber
\end{align}

\noindent In the rest frame of $\phi(\mu_{k})$, its decay rate is given by
\begin{align}
    \Gamma_{Z} &= 
    \frac{\text{sin}^{4}\alpha \, g_{Z}^{2}c_{f}}{8\mu_{k}} \,
    [ \rho(\mu_{k})\Delta \mu^{2} ]
    \int
    % _{\mu_{0}}^{\mu_{1}} 
    \frac{d\mu_{l}^{2}}{2\pi}
    \rho(\mu_{l})
    \int d\Pi_{3}
    \,\, 
    \Gamma_{\mu_{l}x_{f}x_{\bar{f}}}\,.
\end{align}
Here the term in the square brackets is the value of the DM field profile on the SM brane, as approximated by Eq.~\leqn{limita}, and $\Delta\mu^2=2 \mu \Delta\mu $. In the numerical simulations, we set $\Delta\mu=1$ GeV, which is an order-of-magnitude approximation of the spacing of the tower of states in the benchmark 5D model defined in Eq.~(\ref{BM_model}). The coefficient $c_{f}$ is defined as in Eq.~(\ref{eq:cq_def}), except that the sum now includes all kinematically accessible SM fermions. 

% The coefficient $c_{f}$ is the square of the coupling of the $Z$ boson to SM fermions:
% \begin{align}
%     c_{f} = 
%     \frac{g^{2}}{2\text{cos}^{2}\theta_{w}} 
%     \sum_{i}(T^{3}_{i} - \text{sin}^{2}\theta_{w} Q_{i})^{2}\,,
% \end{align}
% where $i$ runs over all kinematically accessible SM fermions. 

The 3-body phase space integral is given by (defining $R\equiv \mu_{l}^{2}/\mu_{k}^{2}$)
\begin{align}
    \int d\Pi_{3} = 
    \frac{\mu_{k}^{2}}{128\pi^{3}} 
    \int_{0}^{1-R} dx_{f} 
    \int_{1-R-x_{f}}^{\frac{1-R-x_{f}}{1-x_{f}}} dx_{\bar{f}},
\end{align} 
and the differential decay rate takes the form
\begin{align}
    \Gamma_{\mu_{l}x_{f}x_{\bar{f}}} = 
    8\mu_{k}^{4}
    \frac{1+R+x_{f}x_{\bar{f}}-x_{f}-x_{\bar{f}}} 
         {|\mu_{k}^{2}(x_{f}+x_{\bar{f}}+R-1) - m_{Z}^{2} + i\Gamma_{Z}m_{Z}|^{2}}\,,
\end{align}
where $x_{f,\bar{f}}\equiv2E_{f,\bar{f}}/\mu_{k}$ are the energy fractions of the fermions, which are assumed to be massless. Note that the equations above allow for a unified treatment of on- and off-shell $Z$ bosons, which turns out to be numerically feasible in this case. 
Once again, we use {\tt Vegas} to perform the integrals and sample from the distributions in the decay product phase space. In general, each produced DM state undergoes a cascade decay, with 
the mass of the remaining DM state getting closer to the gap scale with each step. Correspondingly, typical energies of the produced fermions get smaller at each step of the cascade, and the DM state lifetime gets longer. Eventually, the cascade is effectively terminated when either the fermions become too soft to be detected, or the DM state is too long-lived to decay inside the detector.   

\label{subsubsec:ZDecay}

% \subsubsection{KK Graviton Decay}
% \label{subsubsec:GravDecay}

\section{LHC Bounds}
\label{sec:VisiblePheno}

In this section, we recast an existing LHC analysis to place bounds on the parameter space of near-continuum DM models. We then project the sensitivity reach of the HL-LHC for these models.  

\subsection{Recast}

\begin{figure}
\centering
\includegraphics[width=\textwidth]
{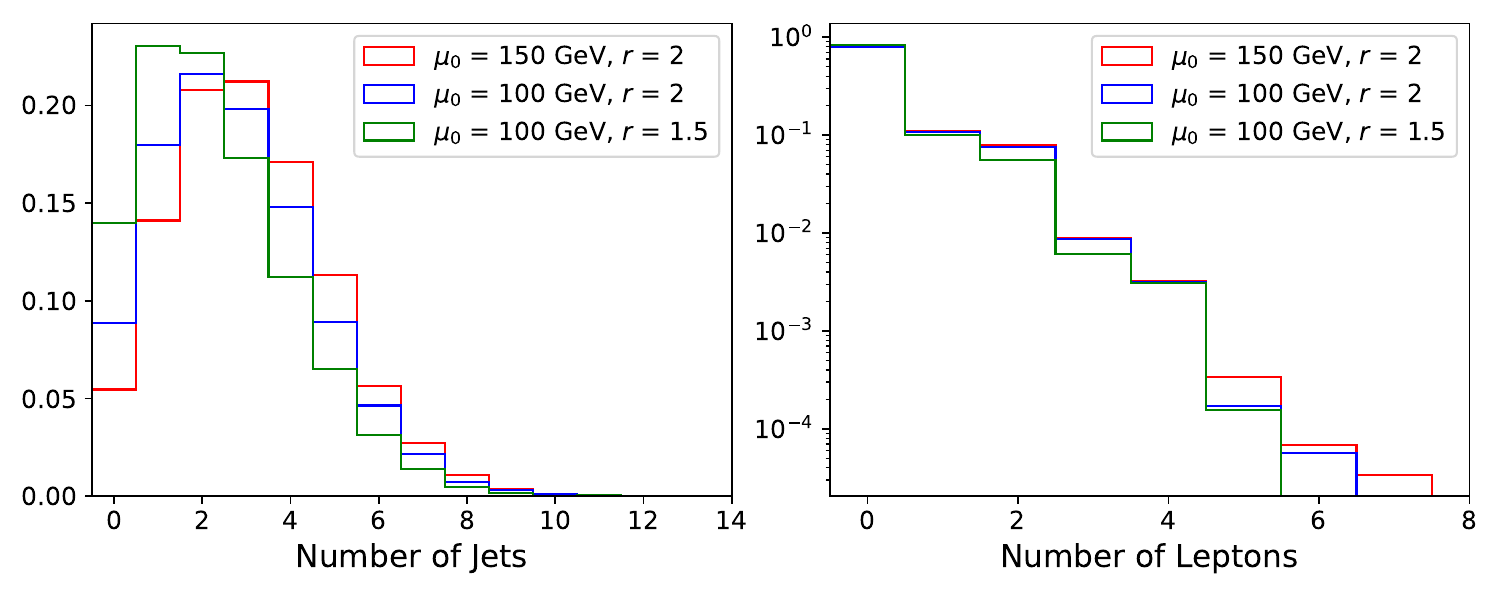}
\caption{Number of final state leptons and jets of the signal for different values of gap scale $\mu_{0}$ and ratio $r\equiv \mu_{p}/\mu_{0}$ after showering and detector simulation. The number of jets and leptons shown here is before the selections for the analysis are applied, but after selecting jets with $p_{T} > $ 30 GeV and $\Delta R < 0.4$, and leptons with $p_{T} > 0.5$ GeV and $\eta < 2.4 \, (2.5)$ for muons (electrons).  }
\label{fig:LepsVSJets}
\end{figure}

The experimental signatures in the analysis chosen for recast should resemble those in our model. Accordingly, we focus on an analysis characterized by high missing energy and a large particle multiplicity. To motivate this choice, it is helpful to first understand how the final states in these models are characterized after parton showering and detector simulation. We examine the structure of the final states in  Fig.~\ref{fig:LepsVSJets}, which shows the multiplicities of jets and charged leptons (electron+muon) in the signal for three benchmark points in the signal space. The histograms are normalized, so the units on the vertical axis are arbitrary. There are slightly more jets and leptons for signal points with a higher gap $\mu_{0}$ and ratio $r$. This is expected since the produced continuum states will on average be heavier, leading on average to more steps in the cascade and therefore more $Z$ decays. 

\begin{figure}[t]
\centering
\includegraphics[width=6in]
{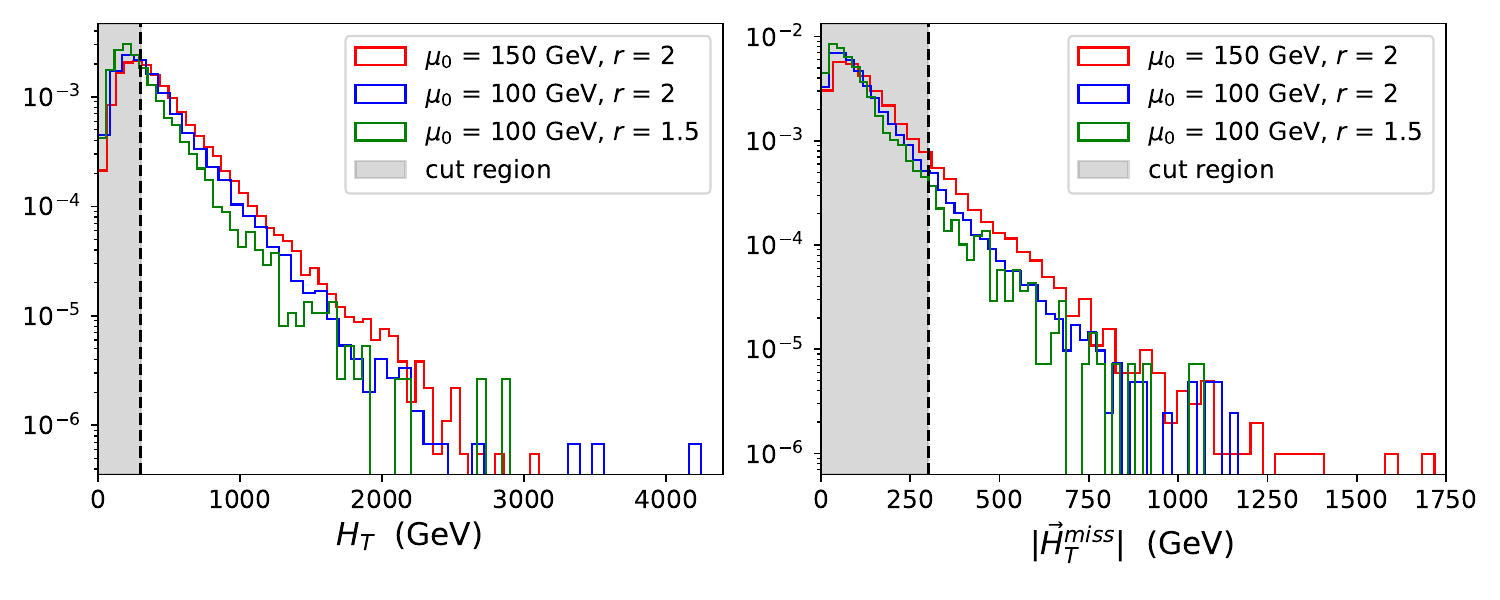}
\caption{Total jet $p_{T}$ ($H_{T}$), and hadronic missing transverse energy ($|\vec{H}_{T}^{\mathrm{miss}}|$) of the signal, shown for various values of the gap scale $\mu_{0}$ and ratio $r\equiv\mu_{p}/\mu_{0}$. The 300 GeV cut for each observable, imposed in the CMS analysis~\cite{Sirunyan_2017}, is shown with a black dashed line. }
\label{fig:ht}
\end{figure}

In the left panel of Fig.~\ref{fig:LepsVSJets}, most events contain between one and six jets, with nearly all events containing at least one jet.  In the right panel, the majority of events have zero leptons. These features are primarily simply dictated by the SM $Z$ branching ratios. This suggests that LHC searches in multijet channels, rather than multilepton, provide the most relevant constraints for our models. One suitable search with both multijet and high missing momentum signatures is performed in Ref.~\cite{Sirunyan_2017} by the CMS collaboration. 
This analysis puts bounds on supersymmetry using Run-2 data with 35.9 fb$^{-1}$. An overview of the selections used in this analysis is as follows: 
\begin{itemize}
    \item $N_\mathrm{jet} \geq 2$  with $|\eta| < 2.4$ for each jet
    \item $H_{T} \equiv \sum_{j} |\vec{p}^{\,j}_{T}| > 300$ GeV, where the sum is over jets with $|\eta| < 2.4$
    \item $|\vec{H}_{T}^{\mathrm{miss}}| \equiv |-\sum_{j} \vec{p}^{\,j}_{T}| > 300$ GeV, where the sum is over jets with $|\eta| < 5$
    \item No identified, isolated electron or muon candidate with $p_{T} > 10 $ GeV 
    \item No isolated track with $m_{T} < 100$ GeV and $p_{T} > 10$ GeV ($p_{T} > 5$ GeV if the track is identified as an electron or muon), where 
    \begin{align}
        m_{T} = \sqrt{2|\vec{p}_{T}^{\,\,\mathrm{tr}}|
        |\vec{p}_{T}^{\,\,\mathrm{miss}}|(1-\cos\Delta\phi_{p_{T}^\mathrm{tr}, p_{T}^\mathrm{miss}})}
    \end{align}
    and $\Delta\phi_{ab}$ is the azimuthal angle between $a$ and $b$.
    \item $\Delta\phi_{H_{T}^{\mathrm{miss}}, j_{1,2}} > 0.5$ and  $\Delta\phi_{H_{T}^\mathrm{miss}, j_{3,4}} > 0.3$, where the jets are ordered in terms of $p_{T}$: $j_{1}, ... j_{N}$, with $j_{1}$ as the highest $p_{T}$ jet.  All jets here must have $|\eta| < 2.4$. 
\end{itemize}   
The near-continuum DM signal events generally have a strong overlap with these selection criteria. In Fig. \ref{fig:ht}, we show two of the kinematic variables used in the selections: the scalar sum $H_{T}$ of the transverse momenta of jets, and the magnitude $|\vec{H}_{T}^{\mathrm{miss}}|$ of the vector sum of the transverse momentum of jets. For larger gap $\mu_{0}$ and ratio $r$, the distributions are skewed to higher values of $H_{T}$ and $|\vec{H}_{T}^{\mathrm{miss}}|$, again reflecting the fact that higher mass continuum states typically result in more steps in the cascade and therefore more jets in each event. 

\begin{figure}[t]
\centering
\includegraphics[width=5in]
{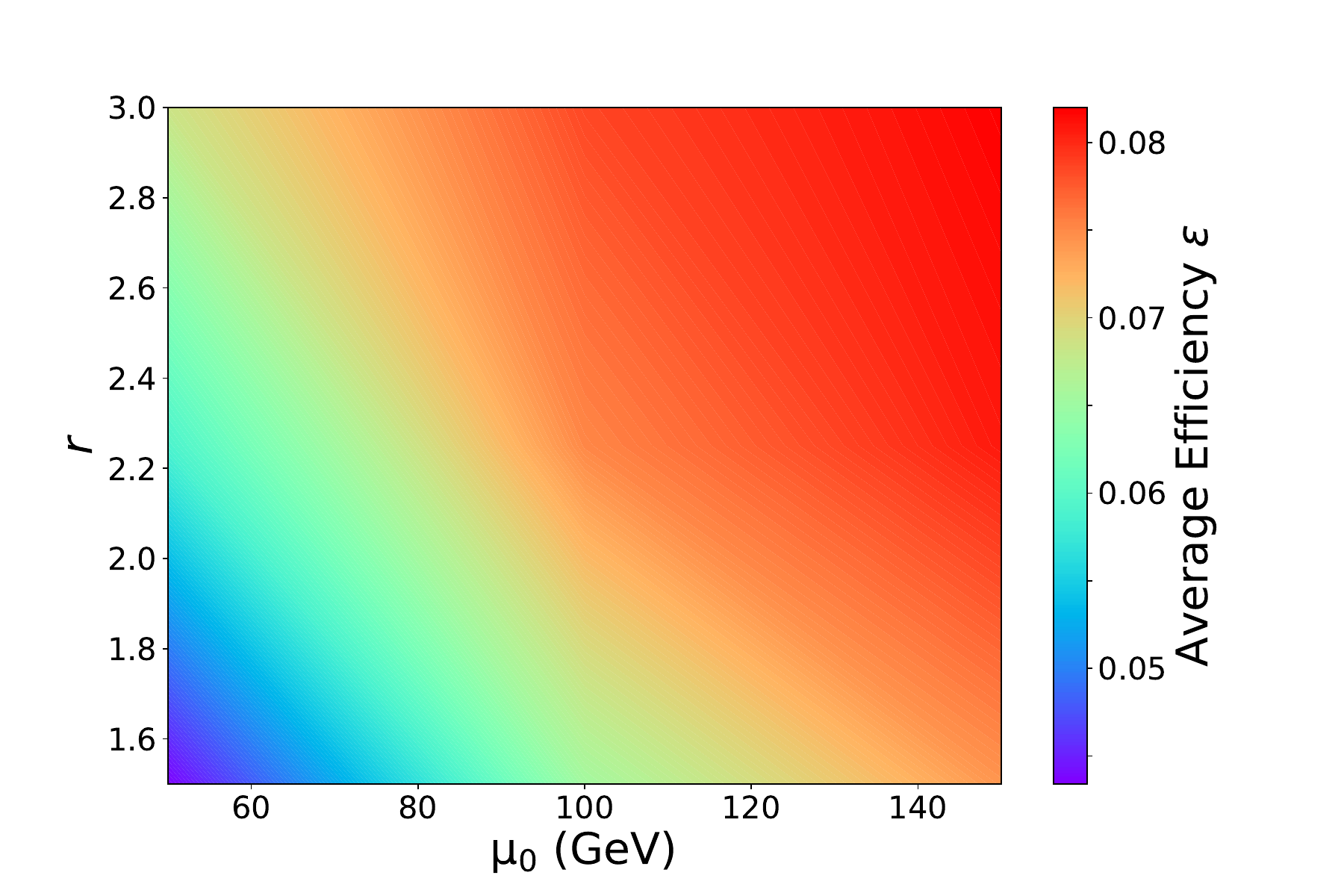}
\caption{Signal efficiencies averaged over the 5 bins in the analysis, as a function of the gap scale $\mu_{0}$ and ratio $r$.}
\label{fig:effs}
\end{figure}

Having established some kinematic features of the signal, we now turn to the details of the recast. Our analysis consists of a difference-of-$\chi^{2}$ test with five bins separated according to the number of jets: $N_\mathrm{jet}$ =  2, $3-4$, $5-6$, $7-8$, $>9$. The $\chi^{2}$ statistic we calculate is 
\begin{align}
    \chi^{2} 
    = 
    \sum_{i=1}^{i=5} 
    \frac{(S_{i}+B_{i}-D_{i})^{2}}{\sigma_{B_{i}}^{2}}
    - 
    \frac{(B_{i}-D_{i})^{2}}{\sigma_{B_{i}}^{2}},
    \label{eq:chisq}
\end{align}
where $S_{i}$, $B_{i}$, and $D_{i}$ are the number of signal, background, and data events respectively in the $i^{\text{th}}$ bin, and $\sigma_{B_{i}}$ is the error on the background in the $i^{\text{th}}$ bin. In Table \ref{tab:tablebins} we show the number of data, background, and signal events for the three benchmark signal points, in each of the five bins.  The numbers of background and data events are taken from the tables in the appendix of~\cite{Sirunyan_2017}. The number of signal events is defined as $S_{i} = \epsilon_{i}\sigma L$, where $\sigma$ is the cross section from Fig.~\ref{fig:XSection} and $L = 35.9$ fb$^{-1}$ is the luminosity used in the CMS analysis. The efficiencies $\epsilon_{i}$ are computed by dividing the number of signal events that pass the selections by the total number of signal events in each bin.  In Fig. \ref{fig:effs}, we average these efficiencies over the five bins, and plot them as a function of gap scale $\mu_{0}$ and ratio $r$. These average efficiencies fall between 4-8\% and grow for larger gap scale $\mu_{0}$ and ratio $r$, which is expected since heavier continuum states are less SM-like. 

%%%%%%%%%%%%%%%%%%%%%%%% Table 1 %%%%%%%%%%%%%%%%%%%
\renewcommand\arraystretch{1.3}
\begin{table}[t]
	\centering
	\begin{tabular}{c|c|c|c|c|c}
		\hline\hline
		& $N_\mathrm{jet} = 2$ & $N_\mathrm{jet} = $ 3,4  & $N_\mathrm{jet} = $ 5,6 & $N_\mathrm{jet} = $ 7,8 & $N_\mathrm{jet} >9$ \\
		\hline
		Data & 48805 & 48269 & 10131 &  1434 &  183  \\
		\hline
		Background & 46471 & 46030 & 10039 & 1537 &  198 \\
		% \hline
            \hline
		$\mu_{0} = 100, r = 1.5$ & 1472 &  1449 & 432 &   90 &  10 \\
            \hline 
            $\mu_{0} = 100, r = 2$ & 833 & 1018 & 369  & 80  & 12 \\
            \hline
            $\mu_{0} = 150, r = 2$ & 627 &  922 &  360 & 80  & 10 \\
		\hline\hline
	\end{tabular}
	\caption{Number of data, background, and signal events for three signal points and in each of the five bins used in this analysis. }
    \label{tab:tablebins}
\end{table}
% \renewcommand\arraystretch{1}
%%%%%%%%%%%%%%%%%%%%%%%%%%%%%%%%%%%%%%%%%%%%%

\begin{figure}[t]
\centering
\includegraphics[width=6in]
{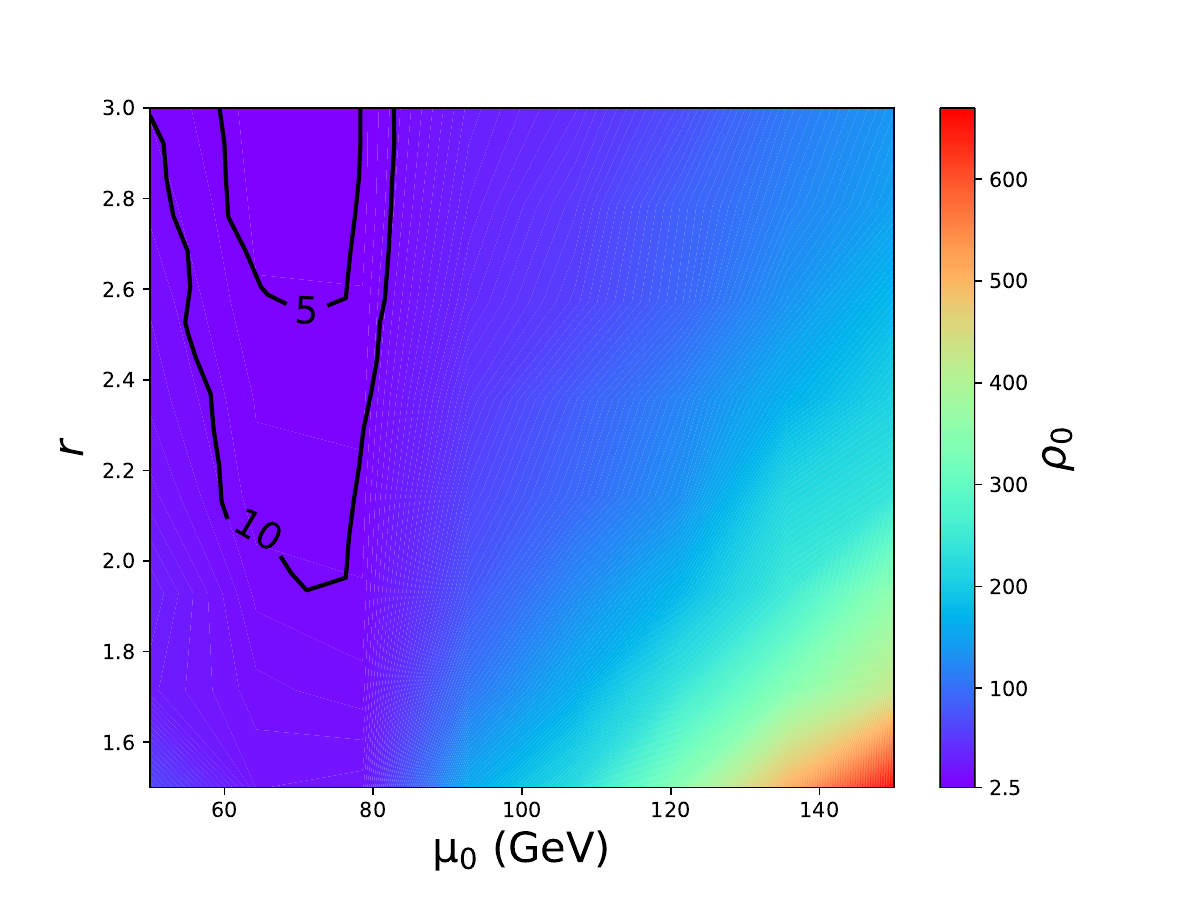}
\caption{Bounds on $\rho_{0}$, as a function of the gap scale $\mu_{0}$ and ratio $r\equiv \mu_{p}/\mu_{0}$, implied by our recast of the CMS search for supersymmetry~\cite{Sirunyan_2017} in the Run-2 LHC data ($L = 35.9$ fb$^{-1}$).}
\label{fig:rho0bounds}
\end{figure}

The cross section $\sigma$ used in the calculation of $S_{i}$ is proportional to $\rho_{0}^{2}\sin^{4}\alpha$. As shown in  \cite{Csaki:2021gfm}, near-continuum DM that reproduces the observed relic abundance implies a relation between $\text{sin}^{2}\alpha$ and $\mu_{0}$, which is plotted in Figure~\ref{fig:dmrelic}. Using this relation, we can place an upper bound on $\rho_{0}$ for fixed gap $\mu_{0}$ and ratio $r$. The critical value of $\chi^{2}$ for five bins at the 95\% confidence level is 9.48. Setting $\chi^2=9.48$ in Eq.~(\ref{eq:chisq}) gives a quadratic equation in $\rho_{0}^{2}$. Since the efficiencies depend on the gap and ratio of the spectral density, the bound on $\rho_{0}$ also depends on $\mu_{0}$ and $r$. These bounds are shown in Fig.~\ref{fig:rho0bounds}. For the gap scale between 60 and 80 GeV, values of $\rho_0$ of order a few can be probed by this analysis. Since, as mentioned above, $\rho_0\sim {\cal O}(1)$ is expected in realistic 5D models, this sensitivity is approaching the level required to probe such models. A modest future increase in sensitivity, which can be achieved with increased statistics and optimized selections in a dedicated search, will allow the LHC to probe a large space of realistic near-continuum DM models.

%For large gap, we see $\mathcal{O}(100)$ bounds on $\rho_{0}$, and for smaller gap the bounds are closer to $\mathcal{O}(10)$. In particular, we draw contours for $\rho_{0}$ = 5 and 10 and observe that some $\mathcal{O}(1)$ bounds are present for a gap scale near the $W$ threshold. 

\subsection{High-Luminosity Projections}
\label{subsubsec:HLProj}

The High-Luminosity (HL) upgrade at the LHC will see up to 3 ab$^{-1}$ of data. We can estimate the sensitivity reach of the search in Ref.~\cite{Sirunyan_2017} at this higher integrated luminosity by rescaling the number of events in the Run 2 analysis. If the luminosity increases by a factor of $N$ ($N\sim$ 83.5 for the current scenario), then the number of expected background events $B$, expected signal events $S$, and statistical error on the background $\sigma_{B}^{\text{stat}}$ scale as follows: 
\begin{align}
    B &\rightarrow NB     \nonumber
    \\
    S &\rightarrow NS
    \\  \nonumber
    \sigma_{B}^{\text{stat}}
    &\rightarrow 
    \sqrt{N}
    \sigma_{B}^{\text{stat}}
    \\   \nonumber
    \sigma_{B}^{\text{sys}}
    &\rightarrow 
    \sqrt{N}
    \sigma_{B}^{\text{sys}}.    
\end{align}
For the systematic uncertainties $\sigma_{B}^{\text{sys}}$, we take the optimistic approach of assuming that they scale in the same way as the statistical uncertainties. For the data $D$, we assume that $D$ is close enough to the $B$ that we can set $D=NB$ in the HL projection. We checked that this gives a very similar bound compared to scaling $D\rightarrow \sqrt{N}D + (N-\sqrt{N})B$, which is the scaling one would choose to keep the $z$-score $\frac{B-D}{\sigma_{B}}$ constant under the rescaling (where $\sigma_{B}$ is the quadratic sum of statistical and systematic uncertainties). We assume that the selection criteria remain the same in the HL-LHC analysis. 

\begin{figure}[t]
\centering
\includegraphics[width=6in]
{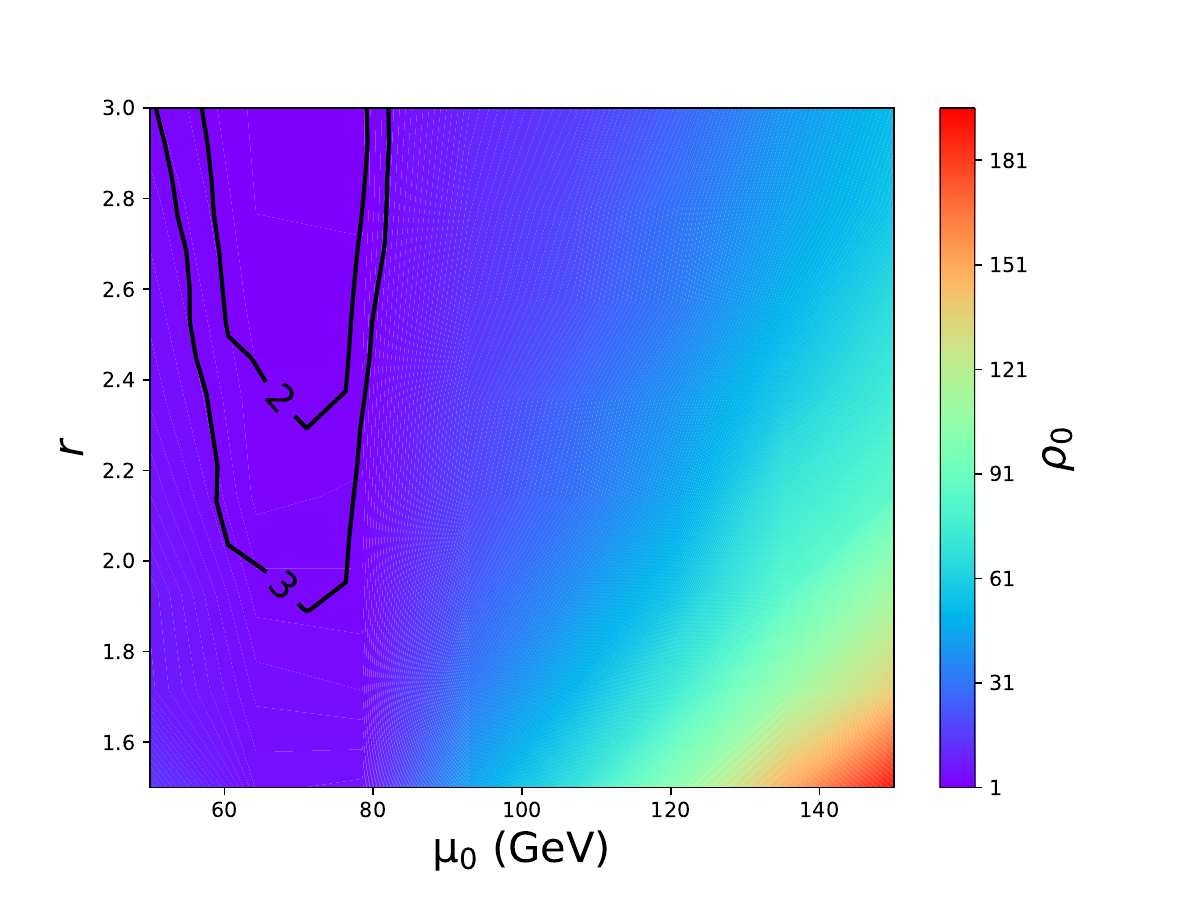}
\caption{Projected sensitivity reach for $\rho_{0}$ as a function of the gap scale $\mu_{0}$ and ratio $r\equiv \mu_{p}/\mu_{0}$ expected at the HL-LHC ($L=3$ ab$^{-1}$).}
\label{fig:HLrho0bounds}
\end{figure}

The extrapolated sensitivity reach obtained for $L=3$ ab$^{-1}$ using the procedure outlined above are shown in Figure~\ref{fig:HLrho0bounds}. The HL-LHC will be sensitive to the near-continuum DM models with realistic $\rho_0\sim \cal{O}(1)$  for gap scales around $50-80$ GeV. Of course, this sensitivity reach can be further improved in a dedicated search for this class of models, where the selection criteria can be optimized for the expected signal. 

\section{Future Lepton Collider Projections}
\label{sec:eeProj}

There are various proposals for electron-positron colliders with center-of-mass energy of $250-500$~GeV and integrated luminosity $\mathcal{O}(\text{ab}^{-1})$, such as the FCC-ee~\cite{FCC:2018evy,Agapov:2022bhm} and the ILC~\cite{ILC:2013jhg,ILCInternationalDevelopmentTeam:2022izu}. The near-continuum DM models pose a promising target for such colliders. The theoretically motivated mass scale in these models of around 100 GeV, within the reach of the next-generation $e^+e^-$ colliders. At the same time, the models are characterized by small production cross sections, due to both the electroweak interaction strength of the $Z$ portal as well as the additional suppression by the mixing angle $\alpha$ and the spread-out spectral density functions. At the LHC, the small signal is concealed under a large SM background, making the search challenging; as we saw in the previous section, the Run-2 searches are only beginning to probe the interesting parameter space, and even the HL-LHC search is expected to leave much of the parameter space unconstrained. In $e^+e^-$ colliders, the expected backgrounds come from electroweak processes and have much smaller cross sections, so that small signals can be identified more easily. In this section, we will estimate the sensitivity reach of the FCC-ee and the ILC for the near-continuum DM model.   

\begin{figure}[t]
\centering
\includegraphics[width=6in]
{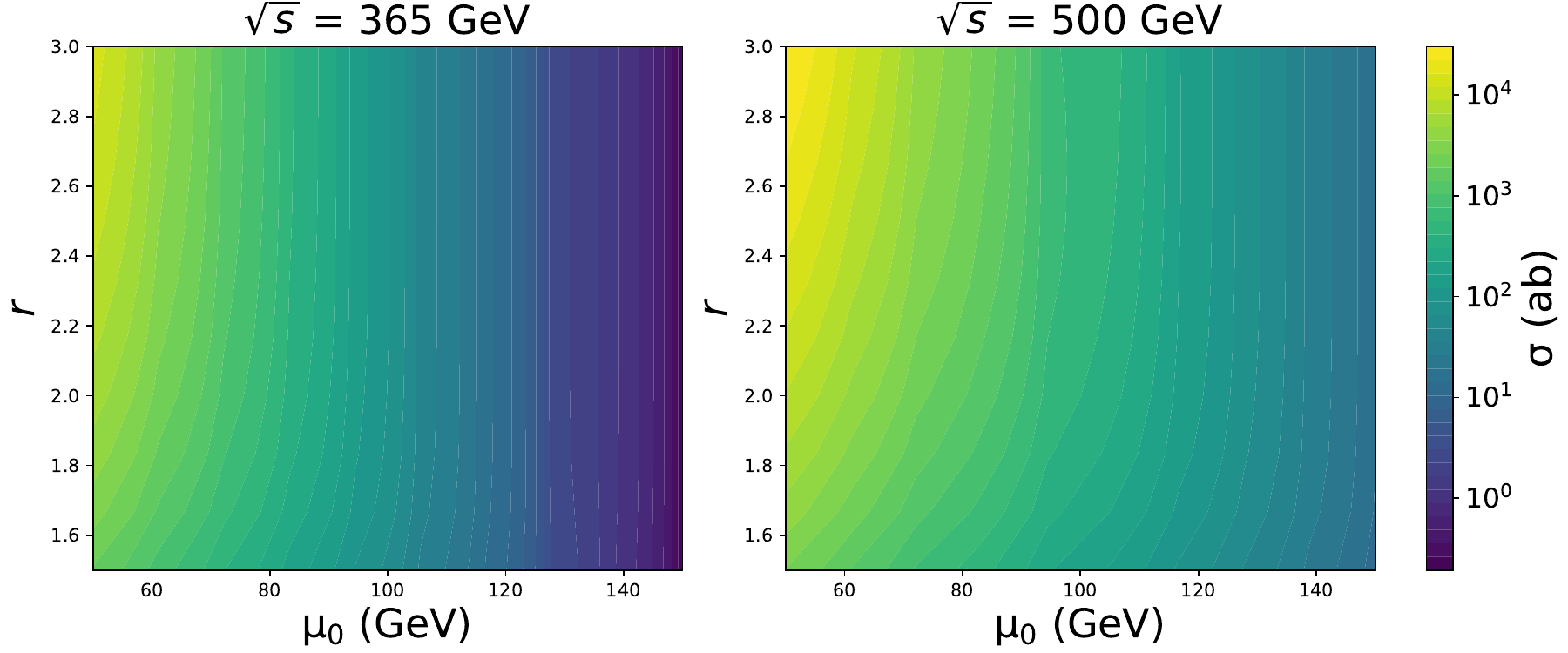}
\caption{Cross sections for pair production of continuum states at an $e^{+}e^{-}$ collider at $\sqrt{s} = 365$ GeV (left) and $\sqrt{s} = 500$ (right), as a function of the gap scale $\mu_{0}$ and ratio $r\equiv \mu_{p}/\mu_{0}$. The overall coefficients $\rho_{0}^{2}$ and $\sin^{4}\alpha$ are set to 1. }
\label{fig:eeCrossSections}
\end{figure}

In Ref.~\cite{Ferrante2023}, cross sections for $e^{+}e^{-}$ were computed with spectral densities computed from the holographic theory. We repeated this calculation with the toy spectral density defined in Eq.~(\ref{eq:ToyRhoDef}). In Figure~\ref{fig:eeCrossSections}, we plot the cross section for pair production of near-continuum DM states in $e^{+}e^{-}$ collisions at $\sqrt{s}=365$ GeV and 500 GeV as a function of the gap scale $\mu_{0}$ and ratio $r$, with $\rho_{0}$ and $\sin^{2}\alpha$ both set to 1 (as in Fig.~\ref{fig:XSection}). In both cases, statistically significant samples of signal events can be produced throughout the relevant parameter space. We have also evaluated the cross section at $\sqrt{s}=250$~GeV. Unfortunately, in that case the cross sections are typically strongly suppressed, since much of the near-continuum DM spectrum remains kinematically inaccessible 
for all but the lowest values of $\mu_0$. Therefore, while the future lepton colliders would collect large amounts of data at 250 GeV to study the Higgs boson, this regime is not optimal for searches for the near-continuum DM, and we do not consider it further.  

We simulate the cascade decays of the near-continuum states as described in section~\ref{subsec:ZDecay} (for further details, see Ref.~\cite{Ferrante2023}.) The resulting parton-level events are stored in the LHE format, and the event files are then passed on to {\tt Pythia}~\cite{Bierlich:2022pfr} for showering and hadronization, and on to {\tt Delphes}~\cite{Selvaggi:2014mya} for a parametrized detector simulation. In the {\tt Delphes} simulation, we use a card that models the IDEA detector, which is a detector concept for the FCC-ee~\cite{IDEAStudyGroup:2025gbt}. In this card we impose a $\Delta R$ cut of 0.4 for the jets and a jet $p_{T}$ minimum of 1 GeV. The leptons are required to have a minimum $p_{T}$ of 0.5 GeV. 

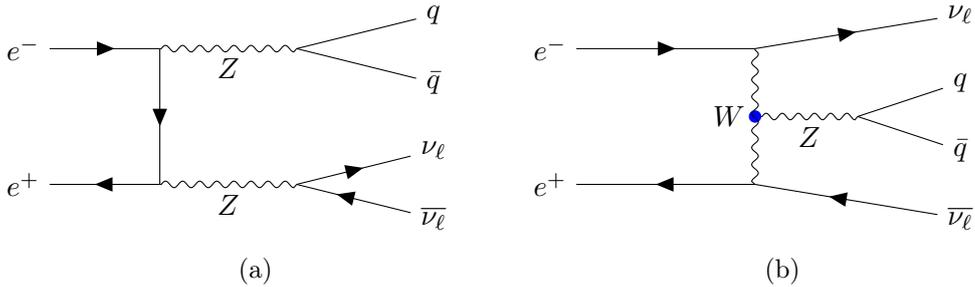
\begin{figure}
\centering
\begin{subfigure}{0.45\textwidth}
  \begin{tikzpicture}
    \begin{feynman}[scale=0.9]
        \vertex (a) at (-2, 1) {$e^-$};
        \vertex (b) at (-2, -1) {$e^+$};
        \vertex (c) at (4, 1.5) {$q$};
        \vertex (d) at (4, 0.5) {$\bar{q}$};
        \vertex (e) at (0, 1);
        \vertex (f) at (0, -1);
        \vertex (g) at (4, -0.5) {$\nu_\ell$};
        \vertex (h) at (4, -1.5) {$\overline{\nu_\ell}$};
        \vertex (i) at (2, 1);
        \vertex (j) at (2, -1);
        
        \diagram* {
            (a) -- [fermion] (e) -- [fermion] (f) -- [fermion] (b),
            (e) -- [boson, edge label'=\(Z\)] (i),
            (f) -- [boson, edge label'=\(Z\)] (j),
            (i) -- [plain] (c),
            (i) -- [plain] (d),
            (j) -- [fermion] (g),
            (j) -- [anti fermion] (h)
        };
    \end{feynman}
\end{tikzpicture}
  \caption{}
  \label{fig:backtobackZ}
\end{subfigure}
\begin{subfigure}{0.45\textwidth}
  \begin{tikzpicture}
    \begin{feynman}[scale=0.9]
        \vertex (a) at (-2, 1) {$e^-$};
        \vertex (b) at (-2, -1) {$e^+$};
        \vertex [dot, blue] (c) at (1, 0) {};
        \vertex [dot, blue] (d) at (2.5, 0);
        \vertex (e) at (1, 1);
        \vertex (f) at (1, -1);
        \vertex (g) at (4, 1.5) {$\nu_\ell$};
        \vertex (h) at (4, -1.5) {$\overline{\nu_\ell}$};
        \vertex (i) at (4, 0.5) {$q$};
        \vertex (j) at (4, -0.5) {$\bar{q}$};
        
        \diagram* {
            (a) -- [fermion] (e) -- [boson, edge label'=\(W\)] (f) -- [fermion] (b),
            (e) -- [fermion] (g),
            (f) -- [anti fermion] (h),
            (c) -- [boson, edge label'=\(Z\)] (d),
            (d) -- [plain] (i),
            (d) -- [plain] (j)
        };
    \end{feynman}
\end{tikzpicture}
  \caption{}
  \label{fig:radiateZ}
\end{subfigure}
\caption{Selected Feynman diagrams for the most relevant background process in lepton colliders, $e^+e^- \rightarrow j j \nu \bar{\nu}$. Both diagrams are dominated by the on-shell $Z$ contributions; however, both on-shell and off-shell $Z$ configurations are included in our background simulation.}
\label{fig:bkgdiagrams}
\end{figure}

As at the LHC, the events with near-continuum DM production in lepton collisions are characterized by large missing energy (ME) and high multiplicity, primarily in jets. Thus, we focus on the jets+ME final state as the signature of this model. The most relevant irreducible background to this search is $e^+e^-\to 2j + \text{ME}$, where the missing energy is due to neutrinos. To estimate the projected sensitivity reach, we simulate the process $e^{+} e^{-} \rightarrow  j j \nu \bar{\nu}$, with selected diagrams shown in Figure \ref{fig:bkgdiagrams}. This simulation is performed using {\tt Madgraph} \cite{Alwall_2011} for parton-level events, followed by {\tt Pythia} and {\tt Delphes}. At 500 GeV this process has a total cross section (with acceptance cuts described above) of $\sigma_{\rm bg} \sim 0.4$~pb. 

\begin{figure}[t]
\centering
\begin{subfigure}{\textwidth}
    \includegraphics[width=6in]{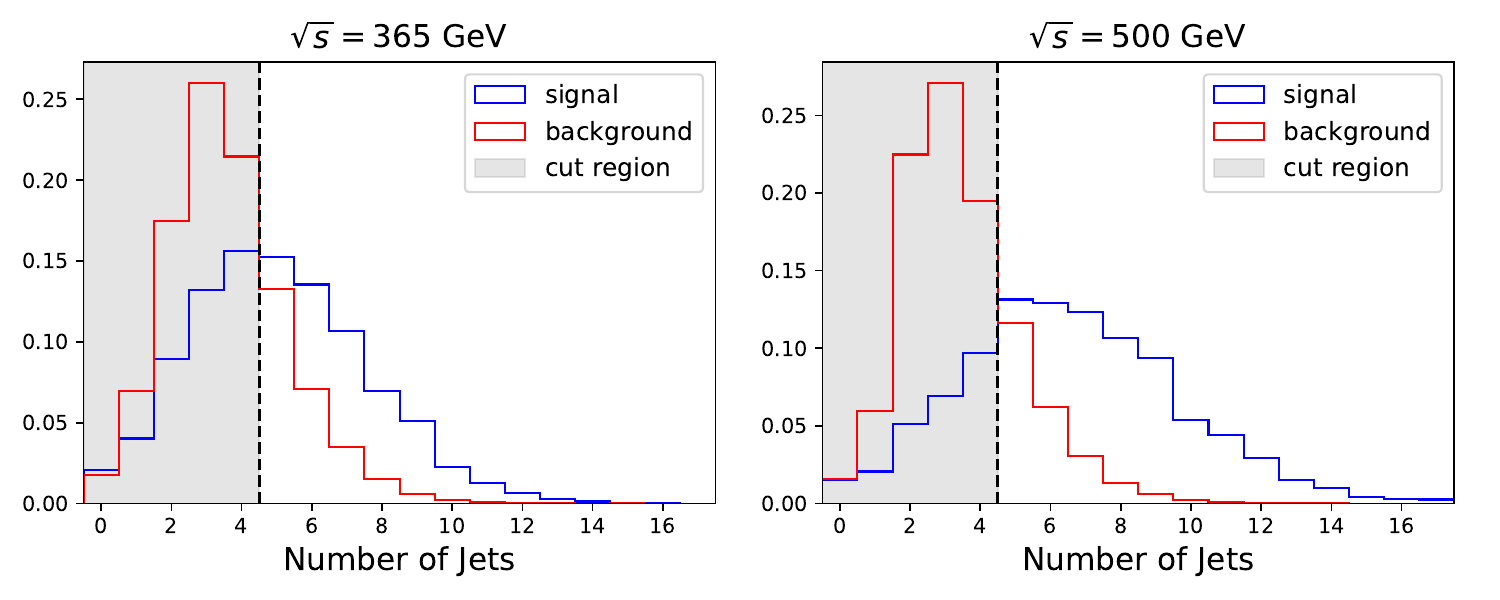}
    \caption{}
    \label{fig:eenJcuts}
\end{subfigure}
\bigskip
\begin{subfigure}{\textwidth}
    \includegraphics[width=6in]{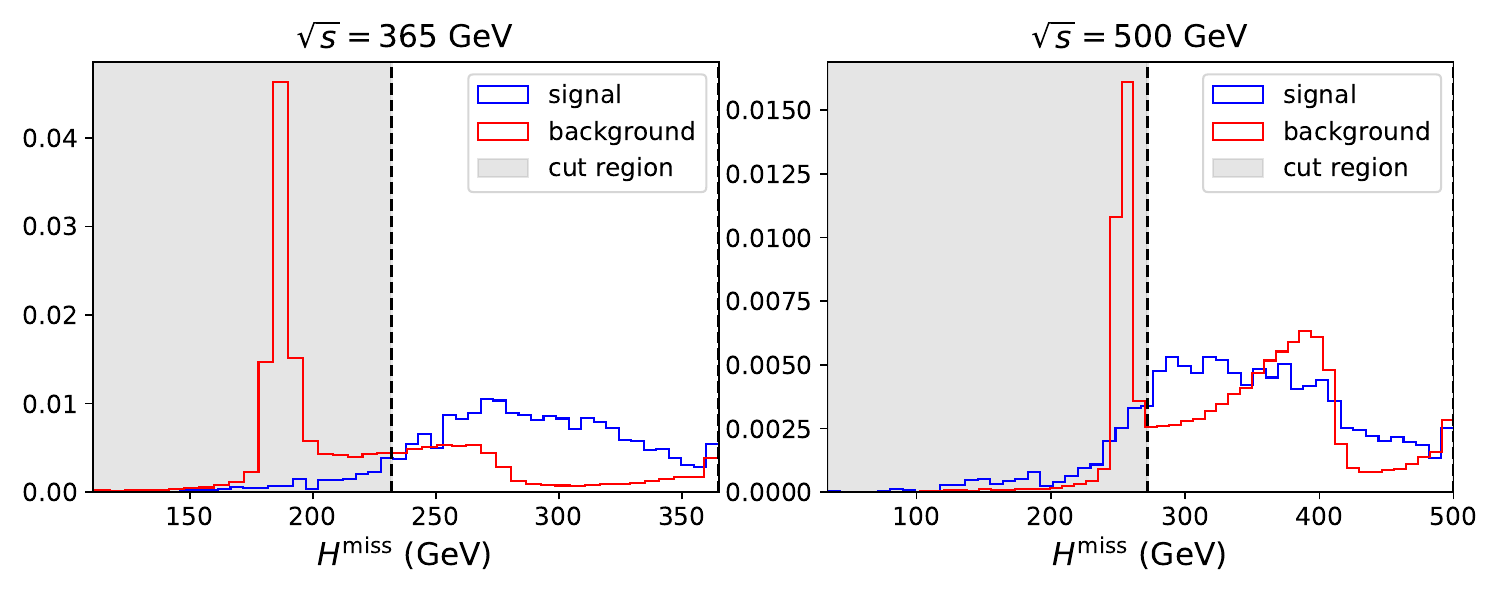}
    \caption{}
    \label{fig:eeMEcuts}
\end{subfigure}
\caption{The signal (blue) and background (red) event distributions in the variables (a) $N_\mathrm{jet}$ and (b) $H^\mathrm{miss}$, in electron-positron collisions with $\sqrt{s}=365$~GeV (left) and $\sqrt{s}=500$~GeV (right). }
\label{fig:eecuts}
\end{figure}

% event selection criteria
We apply the following generic selection criteria to the simulated background and signal events:
\begin{itemize}
    \item $N_\mathrm{jet} \geq 2$, selecting only relevant background events;
    \item $H^{\mathrm{miss}} \equiv \sqrt{s} - |\sum_{j, \ell} \vec{p}^{j, \ell}| < \sqrt{s}$, eliminating signal events that are not triggered on (no DM decays inside the detector).
\end{itemize}
In addition to these global criteria, we also apply analysis cuts customized for each gap scale $\mu_0$ and ratio $r$ for which event samples were simulated. These cuts are chosen to select regions in the variables $N_\mathrm{jet}$ and $H^{\mathrm{miss}}$ where the signal events are most frequent relative to the background. Figures~\ref{fig:eenJcuts} and~\ref{fig:eeMEcuts} show the distribution of signal and background events over $N_\mathrm{jet}$ and $H^{\mathrm{miss}}$ respectively, for near-continuum DM spectra with the gap scale $\mu_0 = 100$ GeV and ratio $r = 2$. The vertical scale is arbitrary. In the case of missing energy in Figure~\ref{fig:eeMEcuts}, the selection cuts are chosen to remove the peaks in the background event distributions associated with the on-shell $Z \rightarrow \nu_\ell \bar{\nu_\ell}$ decays. The peaks appear at $H^{\mathrm{miss}} \sim \sqrt{s}/2$ due to the back-to-back $Z$ production in the $e^+e^-\to ZZ$ process (see Figure \ref{fig:backtobackZ}). The events at the $H^{\mathrm{miss}} \sim \sqrt{s}-m_Z$ peaks are left in the analysis as they have significant overlap with signal regions. These are associated with an on-shell $Z \rightarrow q \bar{q}$ in Figure \ref{fig:radiateZ}.

\begin{figure}[t]
\centering
\includegraphics[width=6in]
{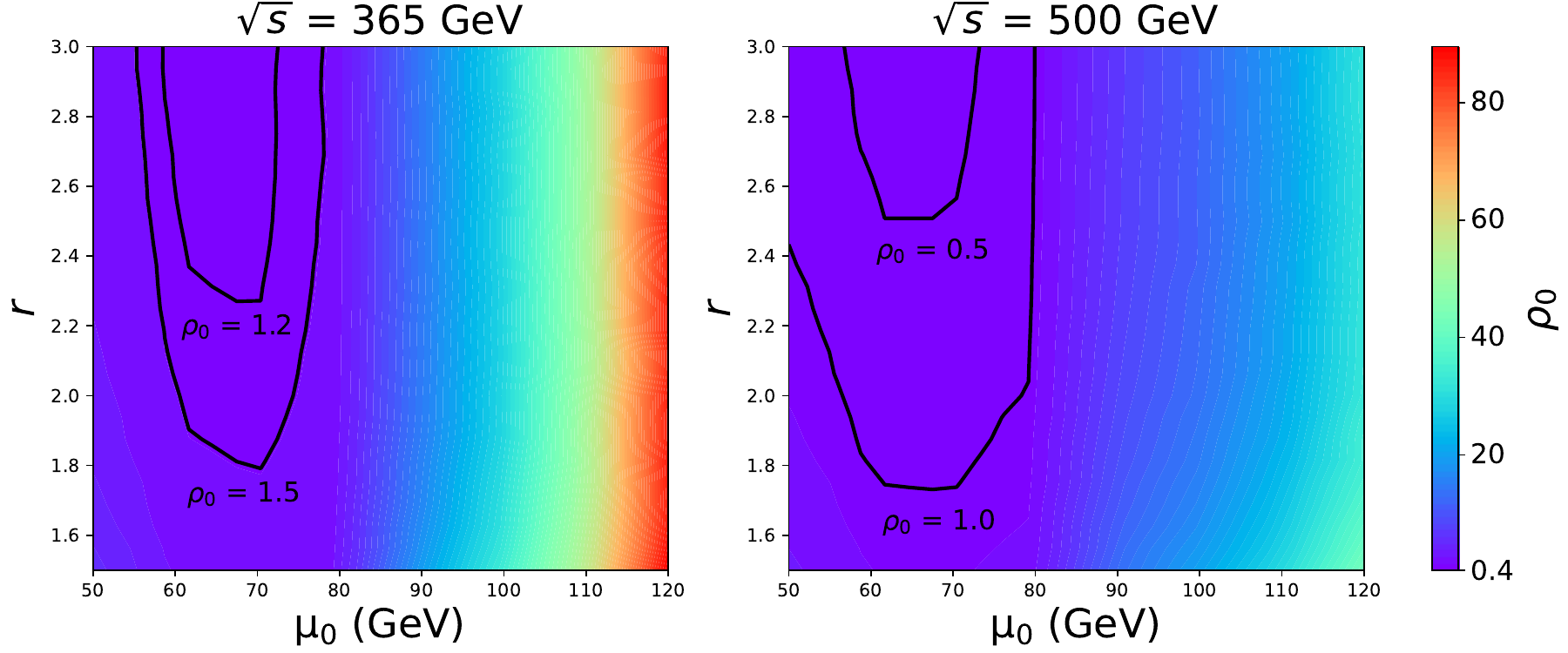}
\caption{Projected bounds on $\rho_{0}$ as a function of gap $\mu_{0}$ and ratio $r\equiv \mu_{p}/\mu_{0}$. Projections are shown for FCC-ee with $\sqrt{s}=365$ GeV and ILC with $\sqrt{s}=500$ GeV, with selection criteria defined in the main text.}
\label{fig:eeProjectionscut}
\end{figure}

In Figure~\ref{fig:eeProjectionscut}, we plot the estimated exclusion reach of the future $e^+e^-$ colliders in terms of  $\rho_{0}$, as a function of the gap scale $\mu_{0}$ and ratio $r$. The bound is calculated by requiring the statistical significance $S / \sqrt{B} \geq 3$. Systematic errors are assumed to be subdominant. The bound on $\rho_{0}$ is determined from the signal and background efficiencies $\epsilon_{S}$ and $\epsilon_{B}$, the background cross section $\sigma_{B}$, and the integrated luminosity $L$ as follows: 
\begin{align}
    \rho_{0 \mathrm{, bound}} = \sqrt{ 
    3 \times \frac{1}{\epsilon_{S}\sigma_{S}}\sqrt{\frac{\epsilon_{B}\sigma_{B}}{L}}
     }\,.
\end{align}
In the plots, we assumed the same integrated luminosity, $L=3$ ab$^{-1}$, at both the FCC-ee ($\sqrt{s}=365$~GeV) and the ILC ($\sqrt{s}=500$~GeV).

As expected, our analysis shows that future lepton colliders have significantly a higher sensitivity reach to the near-continuum DM model compared to the LHC, even with the full HL-LHC dataset. The center-of-mass energy of the $e^+e^-$ collisions plays a very important role in determining the reach, especially for higher values of the gap scale $\mu_0$. This is not surprising, since more of the near-continuum DM modes are kinematically accessible at a higher $\sqrt{s}$. In general, the future lepton colliders have sufficient sensitivity to probe deeply into the interesting regions of the model parameter space for gap scales below 80 GeV.  Above $\mu_0 \sim 80$ GeV, the sensitivity reach is lowered due to the relation of gap scale $\mu_0$ and mixing angle $\sin^2 \alpha$ from DM relic abundance constraints, shown in Figure~\ref{fig:dmrelic}. Improving the reach in that region requires either improvements in the search strategies beyond the very simple cut-and-count analysis considered here, or going to higher center-of-mass energies to enhance the signal rates.

%\begin{figure}[t]
%\centering
%\includegraphics[width=6in]{eeProjections.pdf}
%\caption{Projected bounds on $\rho_{0}$ as a function of gap %$\mu_{0}$ and ratio $r\equiv \mu_{p}/\mu_{0}$. Projections %are shown for FCC-ee with $\sqrt{s}=365$ GeV and ILC with %$\sqrt{s}=500$ GeV.}
%\label{fig:eeProjections}
%\end{figure}

% \begin{figure}[t]
% \centering
% \includegraphics[width=6in]
% {eeProjections_ME_nJ_cut.pdf}
% \caption{Projected bounds on $\rho_{0}$ as a function of gap $\mu_{0}$ and ratio $r\equiv \mu_{p}/\mu_{0}$. Projections are shown for FCC-ee with $\sqrt{s}=365$ GeV and ILC with $\sqrt{s}=500$ GeV. WITH MISSING ENERGY AND N\_JETS CUTS}
% \label{fig:eeProjectionscut}
% \end{figure}

\section{Conclusions and Outlook}
\label{sec:conc}

We have presented the first study of collider constraints on near-continuum dark matter in the $Z$-portal framework, focusing on inclusive signatures at the LHC and projections for future colliders. Using the dedicated Monte Carlo tools developed in Ref.~\cite{Ferrante2023}, which extend standard event generators to handle near-continuous spectra, we simulated production and cascade decays of near-continuum DM states, and performed a recast of the CMS multijet+$H_T^{\mathrm{miss}}$ analysis. Our results translate the Run-2 data into limits on the spectral-density normalization $\rho_0$ as a function of the gap scale $\mu_0$ and shape parameter $r$, showing that the current LHC sensitivity is approaching the theoretically motivated target $\rho\sim {\cal O}(1)$ in some of the parameter space. We also provided High-Luminosity LHC extrapolations of the CMS multijet+$H_T^{\mathrm{miss}}$ search, which show improved sensitivity due to larger statistics. Of course, further improvements in the LHC sensitivity are possible if a dedicated search with selection cuts optimized for the near-continuum DM signal is performed. We hope that our work will motivate the LHC collaborations to pursue such a search. 

Our results highlight the distinctive collider phenomenology of near-continuum dark sectors, where dense spectral structures and multi-step decays produce high-multiplicity, large-$E_T^{\mathrm{miss}}$ events. The LHC analysis presented here establishes a concrete procedure for connecting collider observables to the spectral properties of soft-wall–like geometries, providing a framework applicable to a broad class of continuum dark-sector models.

Looking into the future, lepton colliders offer an especially promising frontier in searches for near-continuum DM. Their precisely controlled electroweak initial states and clean experimental environment allow detailed reconstruction of the final state particles and provide significantly enhanced sensitivity. Even at moderate center-of-mass energies, facilities such as the ILC and FCC-ee can surpass the reach of the LHC, opening a powerful window into dark sectors that interpolate smoothly between particle-like and continuous dynamics.

%%%%%%%%%%%%%%%%%%%%%%%%%%
\section*{Acknowledgments}
We would like to thank Seung J. Lee %, Csaba Csaki, Sungwoo Hong, Gowri Kurup and Wei Xue 
for collaboration and useful discussions on Continuum Dark Matter models. %We are also grateful to Ameen Ismail for helpful discussions.
The authors are supported by the NSF grant PHY-2309456. 
SF is partially supported by the Boochever Fellowship at Cornell University.
LL is also supported by the NSERC Fellowship PGS D - 578021 - 2023.
TY is also funded by  the Samsung Science and Technology Foundation under Project Number SSTF-BA2201-06. 
This work was performed in part at Aspen Center for Physics, which is supported by NSF grant PHY-2210452 and a grant from the Simons Foundation (1161654, Troyer).

\appendix
\label{appendix}

\bibliographystyle{utphys}
\bibliography{ref}

\end{document}